\documentclass[aps,pra,reprint,superscriptaddress,floatfix,longbibliography,nofootinbib]{revtex4-2}
\usepackage{amsmath,amssymb,bm,graphicx,microtype,booktabs,array}
\usepackage[colorlinks=true,linkcolor=red,citecolor=blue,urlcolor=blue]{hyperref}
\usepackage{xcolor}
\graphicspath{{figures/}}
\newcommand{\DKL}{D_{\rm KL}}
\newcommand{\Dt}{\widetilde D}
\newcommand{\Reff}{R_{\rm eff}}
\newcommand{\kB}{k_{\rm B}}
\newcommand{\cL}{\mathcal L}

\newcommand{\LS}{\mathcal L_S}
\newcommand{\M}{\mathsf M}
\newcommand{\E}{\mathsf E}
\newcommand{\Q}{\mathsf Q}
\begin{document}
\title{Autonomous Thermal Switching Outperforms\\
the Frozen Optimum with Vanishing Correlations}
\author{Siyi Deng}
\thanks{These authors contributed equally to this work.}
\affiliation{Department of Physics, Nanchang University, Nanchang 330031, China}

\author{Longxing Xu}
\thanks{These authors contributed equally to this work.}
\affiliation{Department of Physics, Nanchang University, Nanchang 330031, China}

\author{Zhuolin Ye}
\thanks{These authors contributed equally to this work.}
\affiliation{School of Physics, Jiangxi Normal University, Nanchang 330022, China}
\author{Xian He}
\affiliation{Department of Physics, Nanchang University, Nanchang 330031, China}
\author{Jianhui Wang}
\email{Contact author: wangjianhui@ncu.edu.cn}
\affiliation{Department of Physics, Nanchang University, Nanchang 330031, China}
\date{October 16, 2026}
\begin{abstract}
Can autonomous thermal switching outperform the best frozen barrier allocation while keeping system--auxiliary correlations arbitrarily small? We establish this relaxation advantage at fixed lowering budget and visible equilibrium, preserving joint detailed balance. On finite reversible networks, optimal fast switching mixes profiles concentrating the budget on single edges. A three-state model gives a strict gain and its finite-rate threshold. Faster switching preserves the gain while correlations vanish throughout relaxation. Collective and continuous models suggest a colloidal test.
\end{abstract}
\maketitle

\section{Introduction}
\label{sec:introduction}
Equilibrium specifies the final state, but not how quickly it is
reached. Basin free energies set equilibrium populations, whereas
transition-state barriers control transport
\cite{Hanggi1990}.
Several routes to faster relaxation are known
\cite{TezaEtAl2026Review,EsslerKrauth2024,MeibohmKlapp2025,KumarBechhoefer2020}.
Protein conformational dynamics reorganizes kinetic pathways
\cite{FrauenfelderEtAl1991,HenzlerWildmanKern2007}, linking
molecular motion to fluctuating single-molecule reaction kinetics
\cite{LuXie1997,EnglishEtAl2006,GopichSzabo2006}.
Fluctuating barriers
\cite{Zwanzig1992,Berezhkovskii1999,Ankerhold1998,TangEtAl2026Nanopores}
and diffusivity \cite{ChechkinEtAl2017} can likewise reshape
transport statistics.

Barrier fluctuations can accelerate escape through resonant
activation \cite{DoeringGadoua1992,MantegnaSpagnolo2000},
while external protocols shorten equilibration
\cite{Martinez2016,Frim2021,LiQuanTu2017,IlkerEtAl2022}
and optimize finite-time thermodynamic performance
\cite{SivakCrooks2012,Ma2020,LiChenSunDong2022,Ye22,LoosEtAl2024,OikawaEtAl2025}.
Initial-state preparation can improve thermal-machine performance
\cite{LinEtAl2022,LiuWang2026}, and coupling geometry can modify
relaxation \cite{TezaYaacobyRaz2023Boundary}.
Externally scheduled barrier changes can also accelerate
finite-time relaxation through noncommuting generators that
redistribute amplitudes among decay modes \cite{WangLu2026}.
Can internal thermal switching outperform every frozen barrier
allocation at the same lowering budget while preserving visible
equilibrium and joint detailed balance?
Can this advantage persist with arbitrarily small transient
system--auxiliary correlations throughout relaxation?

We establish two complementary results for finite reversible networks.
Optimal fast switching mixes profiles concentrating the full budget
on single edges: exponential activation favors concentration, while
switching relieves complementary bottlenecks.
For a fixed device with a strict fast-limit advantage, sufficiently
rapid finite switching preserves the gain while suppressing
system--auxiliary correlations throughout relaxation.
The slow-rate deficit and peak auxiliary departure are respectively
$O(\Gamma^{-1})$ and $O(\Gamma^{-2})$ in the switching rate $\Gamma$.
An exact three-state model establishes the advantage and its
finite-rate threshold; collective and continuous models extend the
mechanism and suggest a static colloidal realization.

\begin{figure}[!t]
\centering
\includegraphics[width=0.9\columnwidth]{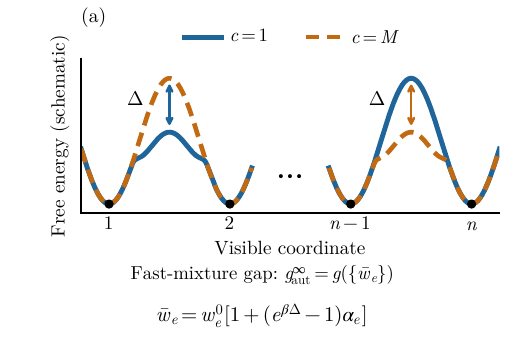}\par
\vspace{2pt}
\includegraphics[width=0.9\columnwidth]{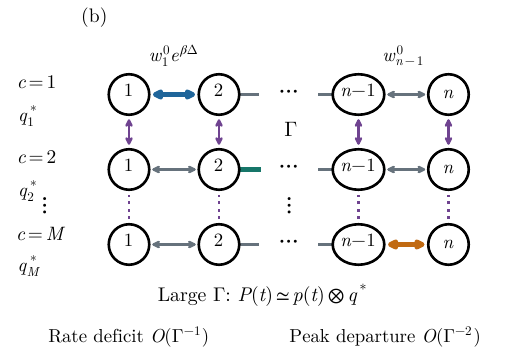}
\caption{\textbf{Reversible model underlying the two main results.}
(a) The curves are conditional profiles at fixed $c$, each spending
$\Delta$ on a different barrier. Fast switching averages their conductances
with weights $q_c^*=\alpha_c$, giving
$g_{\rm aut}^{\infty}=g(\{\bar w_e\})$.
Optimizing the weights yields Eq.~\eqref{eq:general-design};
$g_{\rm st}^{\rm opt}$ is its separately optimized frozen benchmark.
(b) Each row shows visible transitions in one configuration; vertical
links represent reversible switching at scale $\Gamma$.
For this path construction, $M=n-1$ and the joint equilibrium is
$\Pi_{ic}=\pi_iq_c^*$.
At fixed profiles and weights, increasing $\Gamma$ preserves the
averaged conductances while suppressing auxiliary deviations.
The slow-rate deficit is $O(\Gamma^{-1})$ and the peak relative-entropy
departure is $O(\Gamma^{-2})$ under the assumptions of
Eq.~\eqref{eq:hierarchy}.
Ellipses indicate omitted states and configurations; the results
apply to general finite connected reversible networks.}
\label{fig:model}
\end{figure}

\section{Model and main results}
A visible system with states $i=1,\ldots,n$ and equilibrium weights
$\pi_i>0$ is coupled to auxiliary configurations $c=1,\ldots,M$ with
weights $q_c^*>0$. Figure~\ref{fig:model}(a) shows conditional profiles
concentrating a lowering budget $\Delta$ on different edges.
In Fig.~\ref{fig:model}(b), visible transitions connect states within
rows and thermal auxiliary transitions connect rows at switching scale
$\Gamma$. Both obey detailed balance with
$\Pi_{ic}=\pi_iq_c^*$; no external switching schedule is imposed.

Let $g(\mathcal L)$ denote the smallest nonzero eigenvalue of
$-\mathcal L$ for an irreducible reversible generator.
At fixed $\pi$, conductances $\bm w$ determine a generator through
$R_{ij}(\bm w)=w_{\{i,j\}}/\pi_j$, with diagonals fixed by probability
conservation; write $g(\bm w)=g[R(\bm w)]$.
Allowed lowerings give $w_e=w_e^0e^{\beta\delta_e}$, where
$\beta=(\kB T)^{-1}$, $w_e^0>0$, $\delta_e\geq0$, and
$\sum_e\delta_e\leq\Delta$ with $\Delta>0$.
The optimal frozen gap is
$g_{\rm st}^{\rm opt}=\max_{\delta_e\geq0,\,\sum_e\delta_e\leq\Delta}
 g(\{w_e^0e^{\beta\delta_e}\}_e)$.
Let $g_{\rm aut}^{\sup}$ be the supremum of the joint gap over all
admissible auxiliary configurations, equilibrium weights, conditional
profiles, and finite switching rates, holding $\pi$, $\bm w^0$,
and $\Delta$ fixed. With $x=\beta\Delta$, our first result is
\begin{equation}
 g_{\rm aut}^{\sup}
 =\max_{\bm\alpha\in\mathcal S_E}
 g\!\left(
 \left\{w_e^0[1+(e^x-1)\alpha_e]\right\}_e
 \right)
 \geq g_{\rm st}^{\rm opt},
 \label{eq:general-design}
\end{equation}
where $\mathcal S_E=\{\bm\alpha:\alpha_e\geq0,\sum_e\alpha_e=1\}$.
Configuration $e$ puts the full budget on edge $e$ and has weight
$\alpha_e$ [Fig.~\ref{fig:model}(a)]; zero-weight configurations are
omitted. At most one configuration per edge is needed, and the
supremum is approached by fast switching.

For the second result, fix a device and vary only the auxiliary
switching-rate scale $\Gamma$. The auxiliary generators are
$\Gamma\widehat G^{(i)}$, with fixed dimensionless irreducible
$\widehat G^{(i)}$.
Define its joint gap $g_{\rm aut}(\Gamma)=g(\mathcal L_\Gamma)$ and
fast limit $g_{\rm aut}^{\infty}=g(\Reff)$, where
$\Reff=\sum_cq_c^*R^{(c)}$ and $R^{(c)}$ is the visible generator in
configuration $c$. Assume that the averaged gap $g_{\rm aut}^{\infty}$ is simple.
For a fixed full-support preparation $P(0)=p(0)\otimes q^*$, set
$\Dt_{\max}(\Gamma)=\max_{t\geq0}
\DKL[P_\Gamma(t)\Vert p_\Gamma(t)\otimes q^*]$,
where $p_\Gamma$ is the visible marginal and logarithms are natural.
This measures the peak departure from auxiliary equilibrium and
independence. For large $\Gamma$,
\begin{equation}
 \begin{aligned}
  g_{\rm aut}^{\infty}-g_{\rm aut}(\Gamma)
  &=\frac{\mu_2}{\Gamma}+O(\Gamma^{-2}),\\
  \Dt_{\max}(\Gamma)
  &=O(\Gamma^{-2}),
 \end{aligned}
 \label{eq:hierarchy}
\end{equation}
with a device-dependent $\mu_2\geq0$.
Whenever $g_{\rm aut}^{\infty}>g_{\rm st}^{\rm opt}$, sufficiently
large finite switching rates retain a strict advantage with arbitrarily
small departure throughout relaxation.

\section{Reversible joint dynamics}
The joint distribution follows a Markov master equation
\cite{GoutsiasJenkinson2013},
\begin{equation}
 \dot P_{ic}
 =\sum_jR_{ij}^{(c)}P_{jc}
 +\Gamma\sum_{c'}\widehat G_{cc'}^{(i)}P_{ic'},
 \label{eq:joint}
\end{equation}
where the fixed normalized auxiliary generators $\widehat G^{(i)}$
are irreducible and satisfy
$\widehat G_{cc'}^{(i)}q_{c'}^*=\widehat G_{c'c}^{(i)}q_c^*$.
Likewise, $R_{ij}^{(c)}\pi_j=R_{ji}^{(c)}\pi_i$.
Detailed balance gives zero stationary currents and housekeeping entropy
production; perturbed states produce entropy while relaxing
\cite{Schnakenberg1976,Seifert2005,Seifert2012,Bauer2014,FalascoEsposito2021}.
State-dependent $\widehat G^{(i)}$ permits reciprocal kinetic coupling.
The edge conductance is
$w_e^{(c)}=\pi_jR_{ij}^{(c)}=\pi_iR_{ji}^{(c)}$, and
\begin{equation}
 w_e^{(c)}=w_e^0e^{\beta\delta_e^{(c)}},\qquad
 \delta_e^{(c)}\geq0,\qquad
 \sum_e\delta_e^{(c)}=\Delta.
 \label{eq:budget}
\end{equation}
Each conditional profile spends the full budget; frozen profiles may
spend less. Physically, this constraint can describe a fixed coupling
amplitude redistributed among transition regions, with basin free energies
and bare prefactors unchanged. Auxiliary mobility is a separate parameter.

\begin{figure*}[htb]
\centering
\includegraphics[width=\linewidth]{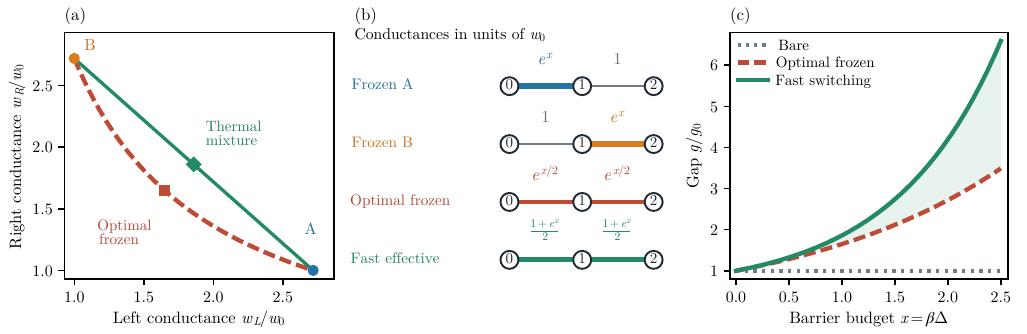}
\caption{\textbf{A minimal analytic witness of strict separation.}
 (a) At fixed $x=\beta\Delta>0$ (shown for $x=1$), full-budget frozen conductances satisfy
 $w_Lw_R=w_0^2e^x$. Equal mixing of the extreme profiles gives a point above
 this curve. (b) Each extreme profile leaves one bottleneck; thermal
 switching balances both. The best frozen allocation splits the lowering
 equally. (c) Gaps normalized by $g_0=3w_0$: the optimal frozen and fast-switching
 values are $e^{x/2}$ and $(1+e^x)/2$, respectively. Their ratio is
 $\cosh(x/2)>1$.}
 \label{fig:resource}
\end{figure*}

\section{Optimal allocation and its physical origin}
With $p_i=\sum_cP_{ic}$, stochastic averaging
\cite{Bo2017,PavliotisStuart2008} gives, for large $\Gamma$ after the initial auxiliary
relaxation,
\begin{equation}
 \dot p=\Reff p+O(\Gamma^{-1}),
 \label{eq:fast}
\end{equation}
where $\Reff=\sum_cq_c^*R^{(c)}$ has conductances
$\bar w_e=\sum_cq_c^*w_e^{(c)}$.
Coarse graining can hide thermodynamic information
\cite{Esposito2012} and induce memory \cite{ZhaoHartichGodec2024,Brandner2025};
finite-rate comparisons therefore use the full joint generator. A detailed derivation of Eq.~\eqref{eq:fast}
is given in Appendix~\ref{app:averaging}.
The reversible gap satisfies
\[
 g(\bm w)=\min_{\substack{\sum_i\pi_i f_i=0\\
                         \sum_i\pi_i f_i^2=1}}
           \sum_{e=\{i,j\}}w_e(f_i-f_j)^2.
\]
It is therefore concave in $\bm w$ and nondecreasing in each conductance
\cite{LevinPeres2017}, as in optimal
network mixing
\cite{Boyd2004,SunBoydXiaoDiaconis2006}.
In particular, a frozen optimum can use the full budget.

Restricting the joint variational principle to auxiliary-independent
trial functions gives $g_{\rm aut}(\Gamma)\leq g(\Reff)$.
Increasing $\Gamma$ adds a nonnegative auxiliary Dirichlet form:
the joint gap is nondecreasing and converges to $g(\Reff)$ as the fast
modes separate. The supremum over rates thus reduces to optimizing
averaged conductances.
For each conditional profile, convexity yields
$e^{\beta\delta_e^{(c)}}\leq
1+(e^{\beta\Delta}-1)\delta_e^{(c)}/\Delta$.
Setting $\alpha_e=\sum_cq_c^*\delta_e^{(c)}/\Delta$ gives
$\bm\alpha\in\mathcal S_E$ and
\[
 \bar w_e\leq w_e^0[1+(e^{\beta\Delta}-1)\alpha_e].
\]
The right-hand side is realized by full-budget single-edge profiles
with weights $\alpha_e$.
Conductance monotonicity makes this construction at least as good as
the original mixture. Every such construction is admissible, and a
full-budget frozen optimum is a one-configuration device, proving
Eq.~\eqref{eq:general-design}. 

Switching can access averaged conductances outside the frozen budget set.
Reproducing these rates with a
frozen profile requires
$\delta_e^{\rm match}=\kB T\ln(\bar w_e/w_e^0)$.
Let $q_e^\ddagger(c)=q_c^*w_e^{(c)}/\bar w_e$ be the auxiliary
distribution conditioned on an equilibrium crossing of edge $e$.
The budget condition gives
\begin{equation}
 \Delta_{\rm match}\equiv\sum_e\delta_e^{\rm match}
 =\Delta+\kB T\sum_e\DKL(q^*\Vert q_e^\ddagger)
 \geq\Delta.
 \label{eq:matching-cost}
\end{equation}
Equality requires every edge lowering to be independent of $c$.
Otherwise, matching the averaged rates costs more frozen budget.
A strict gap advantage additionally requires relief of the relevant
slow-mode bottleneck.
Detailed derivations of Eqs.~\eqref{eq:general-design}
and~\eqref{eq:matching-cost} are given in Appendix~\ref{app:optimization}.

\section{Finite-rate correction and auxiliary departure}
For a fixed device, write $P(t)=p(t)\otimes q^*+Y(t)$.
Auxiliary switching damps $Y$ at rates proportional to $\Gamma$.
Eliminating this fast component changes the slow generator by
$O(\Gamma^{-1})$. Perturbing its simple gap gives the rate deficit
$\mu_2/\Gamma+O(\Gamma^{-2})$; the variational bound
$g_{\rm aut}(\Gamma)\leq g_{\rm aut}^{\infty}$ ensures $\mu_2\geq0$.

The product preparation gives $Y(0)=0$.
Bounded forcing from visible transitions and fast auxiliary damping yield
$\sup_{t\geq0}\|Y(t)\|_1=O(\Gamma^{-1})$.
The auxiliary departure satisfies
\[
 \begin{aligned}
 \Dt(t)
 &\equiv\DKL[P(t)\Vert p(t)\otimes q^*]\\
 &\leq\sum_{i,c}\frac{Y_{ic}(t)^2}{p_i(t)q_c^*},
 \end{aligned}
\]
where the inequality follows from $\ln z\leq z-1$.
Full support of the fixed initial preparation ensures
a positive lower bound on $p_i(t)q_c^*$, uniform in
time and $\Gamma$.
Consequently, the peak departure
$\Dt_{\max}(\Gamma)\equiv\max_{t\geq0}\Dt(t)$
is $O(\Gamma^{-2})$, completing
Eq.~\eqref{eq:hierarchy}.

The different powers in Eq.~\eqref{eq:hierarchy} reflect a first-order
generator correction and locally quadratic relative entropy \cite{Crooks2007}.
Faster switching suppresses departure while preserving the enhanced
averaged conductances.
For the common product preparation, the initial visible relative entropy
fixes the total relaxation entropy production \cite{Ge2010,Sekimoto2010};
switching changes its release rate.
Detailed derivations of Eq.~\eqref{eq:hierarchy}
and the entropy balance are given in Appendix~\ref{app:hierarchy}.
Related speed bounds connect relaxation to dissipation and activity
\cite{ShiraishiFunoSaito2018,FalascoEsposito2020,Gu2023,BaoEtAl2025,VuSaito2023,NagayamaEtAl2025,KolchinskyOhgaIto2024}
and statistical distinguishability
\cite{ItoDechant2020,NicholsonEtAl2020}.

\begin{figure}[!t]
\centering
\includegraphics[width=0.94\linewidth]{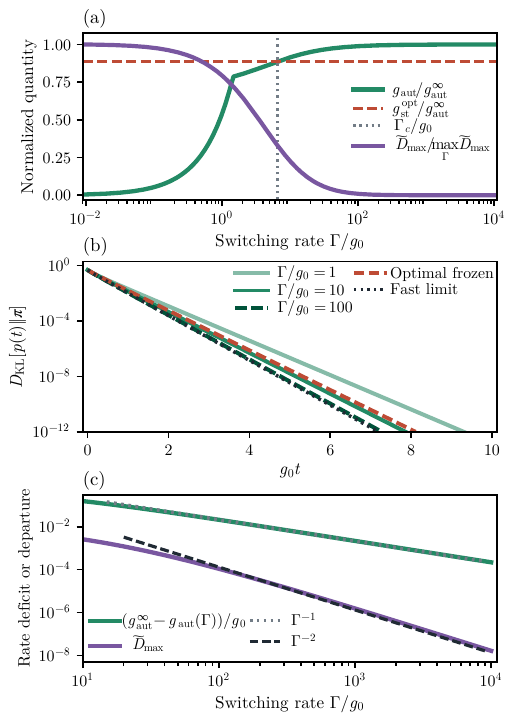}
\caption{\textbf{Finite speed advantage with diminishing auxiliary departure.}
 Exact joint dynamics for $x=1$, $q^*=(1/2,1/2)$, and
 $p(0)=(0.80,0.15,0.05)$; $g_0=3w_0$.
 (a) The full joint gap crosses the best frozen value at
 $\Gamma_c/g_0=4e^{1/2}$ while the normalized peak departure decreases.
 (b) Visible relaxation and the optimal frozen comparison.
 (c) The rate deficit and peak departure approach their inverse and
 inverse-square switching-rate laws.}
 \label{fig:finite}
\end{figure}

\section{Minimal analytic witness}
Consider $0\leftrightarrow1\leftrightarrow2$ with $\pi_i=1/3$, bare
conductances $(w_0,w_0)$, and bare gap $g_0=3w_0$.
For $x>0$, put $a=w_0e^x$ and $b=w_0$.
A full-budget frozen allocation has $uv=ab$ and gap
$g(u,v)=3[u+v-\sqrt{u^2-uv+v^2}]$, maximized by
$u=v=\sqrt{ab}$ [Fig.~\ref{fig:resource}(a)].
Mixing $(a,b)$ and $(b,a)$ instead fixes the averaged conductance sum
at $a+b$. At fixed sum, equal conductances maximize the gap, so equal
mixing attains Eq.~\eqref{eq:general-design}, with
$g_{\rm aut}^{\infty}=g_{\rm aut}^{\sup}=3(a+b)/2$.
Hence
\begin{equation}
 \frac{g_{\rm aut}^{\sup}}{g_{\rm st}^{\rm opt}}
 =\cosh\!\left(\frac{x}{2}\right)>1,
 \label{eq:minimal-witness}
\end{equation}
where $g_{\rm st}^{\rm opt}=3\sqrt{ab}$.
Each extreme profile strongly enhances one edge but leaves
the other as a bottleneck [Fig.~\ref{fig:resource}(b)].
Fast switching averages these complementary profiles,
enhancing both conductances equally.
Exponential activation makes this average exceed the balanced frozen
optimum [Fig.~\ref{fig:resource}(c)]. Matching the averaged rates requires
an excess frozen budget $2\kB T\ln\cosh(x/2)$.
The advantage survives an edgewise lowering cap $d$: for
$\Delta/2<d\leq\Delta$, equally mixing the lowering profiles
$(d,\Delta-d)$ and $(\Delta-d,d)$ gives the fast-limit gap ratio
$\cosh[\beta(d-\Delta/2)]>1$.

For the uncapped device, with auxiliary transition rates $\Gamma/2$
in both directions, the
full six-state spectrum gives $g_{\rm aut}(\Gamma)>g_{\rm st}^{\rm opt}$
if and only if $\Gamma>\Gamma_c$, where
$\Gamma_c=4g_{\rm st}^{\rm opt}$.
Equivalently, the mean switch count during $1/g_{\rm st}^{\rm opt}$
must exceed two.
This exact threshold includes the auxiliary relaxation mode and uses
no fast-switching approximation. Mode crossings can change the slowest
relaxation branch \cite{TezaYaacobyRaz2023Crossing}, so the full joint
gap is essential.
Beyond the crossing in Fig.~\ref{fig:finite}(a), the gap increases
monotonically toward $g_{\rm aut}^{\infty}=g_{\rm aut}^{\sup}$.

We examine Eq.~\eqref{eq:hierarchy} in the same
three-state device by varying only $\Gamma$ at fixed
conditional profiles and equilibrium weights.
All comparisons use the same full-support visible
preparation $p(0)$, with $P(0)=p(0)\otimes q^*$
for the autonomous device.
Figure~\ref{fig:finite}(b) compares the visible
relative entropy $\DKL[p(t)\Vert\pi]$ with the optimal
frozen reference. The $\Gamma/g_0=1$ curve eventually
decays more slowly, whereas the higher-rate curves approach
the fast limit.

In general, the departure decomposes as
$\Dt=I(S{:}C)+\DKL(q\Vert q^*)$, where
$q_c=\sum_iP_{ic}$ and $I(S{:}C)$ is the mutual
information between the visible and auxiliary states
\cite{HorowitzEsposito2014}.
Here state-independent switching and $q(0)=q^*$
ensure $q(t)=q^*$, so $\Dt(t)=I(S{:}C)$ and
$\Dt_{\max}$ is the peak mutual information.
The $\Gamma^{-1}$ gap deficit and $\Gamma^{-2}$ peak correlation
in Fig.~\ref{fig:finite}(c) confirm this separation:
the gap approaches its enhanced fast limit while correlations
vanish uniformly throughout relaxation.
The spectral advantage concerns asymptotic decay
of generic perturbations; it does not require
observable distances to be ordered at every time.
Derivations and numerical validation for the three-state model
are given in Appendix~\ref{app:three-state}.

\begin{figure}[!t]
\centering
\includegraphics[width=\linewidth]{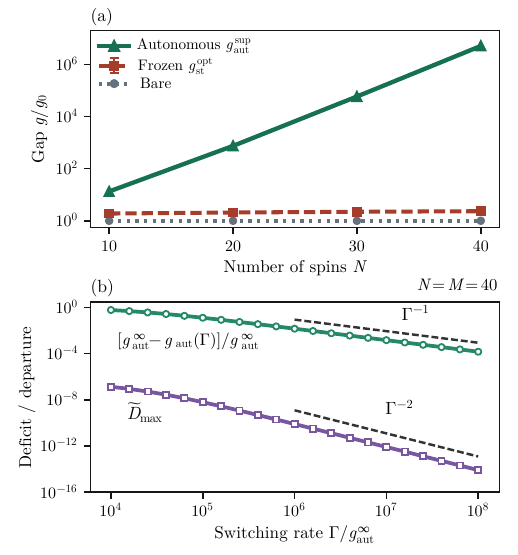}
\caption{\textbf{Optimal collective relaxation and the finite-rate hierarchy.}
(a) Globally optimal autonomous and frozen gaps for the Curie--Weiss
magnetization chain at $\beta J=1.2$ and $h=0$, with identical
budget $\Delta$, equilibrium weights, and bare rates at each $N$.
The autonomous optimum uses $N$ single-edge configurations.
Error bars bracket the frozen optimum to relative width below
$2\times10^{-7}$, unresolved at this scale.
(b) The same optimal device at $N=M=40$: relative joint-gap deficit
(circles) and peak auxiliary departure (squares) versus switching rate.
Dashed lines indicate the inverse and inverse-square laws of
Eq.~\eqref{eq:hierarchy}; profiles, weights, and preparation remain fixed.}
\label{fig:validation}
\end{figure}

\section{Collective and continuous realizations}
For a Curie--Weiss magnet with $N$ spins $s_k=\pm1$,
the energy at magnetization $m=\sum_k s_k$ is
$E_N(m)=-Jm^2/(2N)-hm$, with ferromagnetic coupling $J>0$
and field $h$.
Permutation-symmetric single-spin-flip dynamics reduces exactly
to a reversible path on $m=-N,-N+2,\ldots,N$
\cite{LevinPeres2017}, with
$\pi_N(m)\propto\binom{N}{(N+m)/2}e^{-\beta E_N(m)}$.
At each $N$, equilibrium weights, bare rates, and the total
barrier-lowering budget $\Delta$ are fixed.
Equation~\eqref{eq:general-design} gives the optimal fast-switching
mixture of $N$ configurations, each concentrating the budget on one
magnetization edge. The frozen benchmark optimizes all nonnegative
edge lowerings with sum at most $\Delta$.

For $\beta J=1.2$ and $h=0$, Fig.~\ref{fig:validation}(a) establishes
$g_{\rm aut}^{\sup}>g_{\rm st}^{\rm opt}$ at $N=10,20,30,40$,
with gaps normalized by the bare gap $g_0=g(R^{(0)})$ at each size.
The rapid growth reflects the approximately linear increase of
$\Delta$ with $N$:
uncapped concentration on one edge produces the activation factor
$e^{\beta\Delta}$.
With the optimal $N=40$ device's profiles, weights, and preparation fixed,
varying only $\Gamma$ recovers the hierarchy of
Eq.~\eqref{eq:hierarchy} [Fig.~\ref{fig:validation}(b)].
The gap approaches $g_{\rm aut}^{\infty}=g_{\rm aut}^{\sup}$
while the peak auxiliary departure vanishes.
A simpler two-configuration device already exceeds the global frozen
optimum by $12.9\%$--$15.8\%$ in the fast limit.
Model details, global optimization certificates, and numerical
checks are given in Appendix~\ref{app:collective}.

To connect the mechanism to experiment, we consider a colloidal particle
in a static conservative potential with three longitudinal wells in $x$
and two transverse wells in $y$ \cite{Ciliberto2017}.
Thermal motion in $y$ redistributes a fixed coupling amplitude between
visible transition regions; pinning $y$ defines the frozen family.
Direct Fokker--Planck calculations \cite{Holubec2019,ScharfetterGummel1969}
give a joint gap $41.1\%$ above the best pinned slice for a six-well
potential whose coupling leaves the minimum curvatures unchanged
(see Appendix~\ref{app:continuous}).
The scaled diffusion-rate ratio is
$r=(D_y/\ell_y^2)/(D_x/\ell_x^2)=4$, where $D_x,D_y$ are diffusivities
and $\ell_x,\ell_y$ characteristic lengths.
Each mobile--pinned comparison fixes the bare potential, coupling budget,
and longitudinal mobility.
Basin factorization and visible equilibrium matching are approximate,
so this tests the mechanism beyond the exact finite-state assumptions.

\looseness=-1
Optical landscape shaping and static-trap hopping provide building
blocks for a test \cite{Grier2003,DufresneGrier1998,McCann1999}.
A stiff transverse trap would select a pinned slice; scanning its center
and checking convergence with stiffness would characterize the frozen
family \cite{BergSorensenFlyvbjerg2004}.
Removing this constraint would give the mobile comparison at
unchanged longitudinal potential, coupling, and mobility.
Each landscape would remain fixed during relaxation.
Joint trajectories could resolve slow modes and transient correlations
\cite{Wu2009,FrishmanRonceray2020,SchutteEtAl2011,SchwantesEtAl2014};
visible-only readout can miss auxiliary modes \cite{WuJia2025}.
Equilibrium forward and reverse fluxes would test detailed balance
\cite{RoichmanEtAl2008}.

\section{Conclusions}
Autonomous thermal switching can outperform the best frozen barrier
allocation while preserving joint detailed balance, visible
equilibrium weights, and the lowering budget.
Optimal fast switching in finite reversible networks mixes full-budget
single-edge profiles, combining exponential activation with relief of
complementary bottlenecks.
For a fixed device, the slow-rate deficit is $O(\Gamma^{-1})$ while
the peak auxiliary departure is $O(\Gamma^{-2})$.
A strict fast-limit advantage therefore survives at sufficiently
large finite switching rates with arbitrarily small correlations
throughout relaxation.
The three-state model establishes an exact finite-rate threshold;
collective and continuous models extend the mechanism and suggest
a colloidal test in a static optical landscape.

\begin{acknowledgments}
We are  grateful to Viktor Holubec for his invaluable comments on earlier versions of our manuscript.
This work was supported by the National Natural Science Foundation of China (No. 12465009) and the Major Program of the Jiangxi Provincial Natural Science Foundation (No. 20224ACB201007). Z.Y. acknowledges support from the Jiangxi Provincial Natural Science Foundation (No. 20232BAB211027), the Special Project for Cultivating Early-Career Youth Talents in Science and Technology of Jiangxi Province (No. 20244BCE52030). 
\end{acknowledgments}

\appendix
\section{Averaged dynamics}
\label{app:averaging}
We write $g_{\rm aut}(\Gamma)$ for a fixed device's joint gap,
$g_{\rm aut}^{\infty}=g(\Reff)$ for its fast-switching limit, and
$g_{\rm aut}^{\sup}$ for the supremum over admissible devices and
switching rates. When a decay branch is tracked separately, we denote it
by $r_{\rm vis}(\Gamma)$; it equals the joint gap at sufficiently large
$\Gamma$, but need not do so at smaller rates.

\label{sec:model}
\label{sec:fast}

We derive Eq.~\eqref{eq:fast} from the finite-state model defined by
Eqs.~\eqref{eq:joint} and~\eqref{eq:budget}. The basin free energies are assumed
to be additive:
\begin{equation}
 F_{ic}=\epsilon_i+\varepsilon_c.
 \label{app:eq:additive-minima}
\end{equation}
Thus the equilibrium distribution is \(\Pi_{ic}=\pi_iq_c^*\), with
\(\pi_i=Z_S^{-1}e^{-\beta\epsilon_i}\),
\(q_c^*=Z_C^{-1}e^{-\beta\varepsilon_c}\), and
\(\beta=(\kB T)^{-1}\). All conditional generators are held fixed as
\(\Gamma\) varies. Each auxiliary generator \(\widehat G^{(i)}\)
is assumed irreducible, with stationary distribution \(q^*\).

\subsection{Physical meaning of the lowering budget}
A passive auxiliary can redistribute a fixed coupling amplitude between
two transition regions. For example,
\[
 \begin{gathered}
 U_{\rm int}(x,y)
 =-\Delta\left[h_L(x)s(y)+h_R(x)(1-s(y))\right],\\
 0\leq s(y)\leq1,
 \end{gathered}
\]
with normalized saddle-localized functions \(h_L,h_R\), assigns
complementary coupling coefficients whose sum is \(\Delta\).
Thermal motion of \(y\) changes their allocation without changing the
potential in time. The budget constrains the coupling structure, not
dissipated work or auxiliary mobility. Identifying these coefficients
with barrier lowerings requires localized couplings, fixed basin free
energies, and bare prefactors. Appendix~\ref{app:continuous} retains
the actual saddle geometry and basin weights.

\subsection{Averaged dynamics}
Following the standard stochastic-averaging framework
\cite{PavliotisStuart2008}, we separate the
auxiliary-equilibrated component from fast deviations.
For a joint vector \(X\) and a system vector \(p\),
define the marginalization and embedding maps by
\begin{equation}
 (\M X)_i=\sum_cX_{ic},
 \qquad
 (\E p)_{ic}=p_iq_c^*.
 \label{app:eq:projections}
\end{equation}
These maps satisfy \(\M\E=I\). The projections onto the product
subspace and its complement are \(\mathsf P=\E\M\) and
\(\Q=I-\mathsf P\), respectively. Write the joint generator as
\(\mathcal L_\Gamma=\LS+\Gamma G\),
where \((\LS X)_{ic}=\sum_jR_{ij}^{(c)}X_{jc}\) and
\((GX)_{ic}=\sum_{c'}\widehat G_{cc'}^{(i)}X_{ic'}\).
Probability conservation and \(\widehat G^{(i)}q^*=0\) give
\(\M G=0,\qquad G\E=0,\qquad
\mathsf P G=G\mathsf P=0\).
Irreducibility makes \(G\) invertible on \(\operatorname{ran}\Q\).
Denote this restricted inverse by \(\tilde G\), extended by zero on
\(\operatorname{ran}\mathsf P\).

For fixed positive times as \(\Gamma\to\infty\), the initial fast
transient has decayed, and the joint state admits the expansion
\begin{equation}
 P=\E p+\Gamma^{-1}\chi+O(\Gamma^{-2}),
 \label{app:eq:slow-manifold}
\end{equation}
with \(p=\M P\) and \(\M\chi=0\). Substituting this expansion
into \(\dot P=\mathcal L_\Gamma P\) and applying \(\Q\) at
order \(\Gamma^0\) yields
\(G\chi=-\Q\LS\E p\).
Hence \(\chi=-\tilde G\Q\LS\E p\), which remains bounded as
\(\Gamma\to\infty\). Applying \(\M\) to the exact joint equation
instead gives
\(\dot p=\M\LS P\).
Inserting Eq.~\eqref{app:eq:slow-manifold} therefore gives
\begin{equation}
 \dot p=\Reff p+O(\Gamma^{-1}),
 \label{app:eq:fast-reduction}
\end{equation}
where
\begin{equation}
 \Reff=\M\LS\E=\sum_cq_c^*R^{(c)}.
 \label{app:eq:Reff}
\end{equation}
The last equality follows directly from
Eq.~\eqref{app:eq:projections}.
Equation~\eqref{app:eq:fast-reduction}
reproduces Eq.~\eqref{eq:fast}, with the averaged generator
defined in Eq.~\eqref{app:eq:Reff}.
The residual is therefore $O(\Gamma^{-1})$ after
the initial auxiliary relaxation. For the product
preparation $Y(0)=0$, this estimate holds uniformly
for all $t\geq0$.


\section{Optimal autonomous design and matching cost}
\label{app:optimization}
\label{sec:optimization}

Throughout, the visible graph is finite and connected, and the
equilibrium weights \(\pi_i>0\) and bare conductances \(w_e^0>0\)
are fixed. We first relate the joint finite-rate gap to the averaged gap,
then optimize the latter over conditional barrier profiles.
Finally, we derive the frozen matching cost in Eq.~\eqref{eq:matching-cost}.

\subsection{Variational formulation}
Let \(x=\beta\Delta\), with \(\Delta>0\), and define
\begin{equation}
 \mathcal A_\Delta
 =\left\{\boldsymbol\delta:
 \delta_e\geq0,\ \sum_e\delta_e\leq\Delta\right\}.
 \label{app:eq:static-class}
\end{equation}
The conditional profiles in the main model spend exactly $\Delta$.
Allowing less in $\mathcal A_\Delta$ only enlarges the class used for
an upper bound; the construction below spends the full budget.
For fixed \(\pi\), the reversible gap has the variational representation
\begin{equation}
 g(\boldsymbol w)
 =\min_{\substack{\sum_i\pi_i f_i=0\\
                  \sum_i\pi_i f_i^2=1}}
 \sum_{e=\{i,j\}}w_e(f_i-f_j)^2.
 \label{app:eq:gap-variational}
\end{equation}
Each trial expression is linear in the conductances. Their minimum is
therefore concave and nondecreasing in every \(w_e\).

\subsection{Finite-rate bound and convergence}
We keep \(\LS\) and \(G\) fixed while varying \(\Gamma\).
Each conditional auxiliary generator is assumed irreducible with
stationary distribution \(q^*\), and \(\Reff\) is irreducible.
The joint equilibrium \(\Pi_{ic}=\pi_iq_c^*\) is independent of
\(\Gamma\). With \(D_\Pi=\operatorname{diag}(\Pi_{ic})\),
the symmetrized generator is
\begin{equation}
 \cL_\Gamma^{\rm sym}
 =D_\Pi^{-1/2}\cL_\Gamma D_\Pi^{1/2}
 =\LS^{\rm sym}+\Gamma G_{\rm sym}.
 \label{app:eq:symmetrization}
\end{equation}
Detailed balance makes both terms real symmetric, and
\(G_{\rm sym}\preceq0\). Thus, for \(\Gamma_2>\Gamma_1\),
\begin{equation}
 \cL_{\Gamma_2}^{\rm sym}-\cL_{\Gamma_1}^{\rm sym}
 =(\Gamma_2-\Gamma_1)G_{\rm sym}\preceq0.
 \label{app:eq:Loewner-order}
\end{equation}
Ordering the eigenvalues as
\(0=\lambda_1(\Gamma)\geq\lambda_2(\Gamma)\geq\cdots\),
the min--max principle gives
\begin{equation}
 \lambda_k(\Gamma_2)\leq\lambda_k(\Gamma_1),
 \qquad k\geq2.
 \label{app:eq:eigenvalue-ordering}
\end{equation}
Hence the joint gap \(g_{\rm aut}(\Gamma)=-\lambda_2(\Gamma)\)
is nondecreasing.

Restricting the joint Dirichlet trial function to \(F_{ic}=f_i\)
eliminates the auxiliary contribution and leaves the Dirichlet form
of \(\Reff\). Minimization over this restricted set gives
\begin{equation}
 g_{\rm aut}(\Gamma)\leq g(\Reff).
 \label{app:eq:finite-upper}
\end{equation}
To establish the limit, let \(n\) be the number of visible states and
define \(T_{(i,c),j}=\delta_{ij}\sqrt{q_c^*}\).
The columns of \(T\) form an orthonormal basis of
\(\ker G_{\rm sym}\). With \(D_\pi=\operatorname{diag}(\pi_i)\),
projection onto this subspace gives
\begin{equation}
 T^{\mathsf T}\LS^{\rm sym}T
 =D_\pi^{-1/2}\Reff D_\pi^{1/2}.
 \label{app:eq:projected-slow-generator}
\end{equation}
For \(\Gamma^{-1}\cL_\Gamma^{\rm sym}
=G_{\rm sym}+\Gamma^{-1}\LS^{\rm sym}\), finite-dimensional
perturbation theory splits the \(n\)-dimensional zero eigenspace
according to this projected matrix. Consequently, the \(n\) finite
eigenvalues satisfy
\begin{equation}
 \lambda_k(\cL_\Gamma)
 =\lambda_k(\Reff)+O(\Gamma^{-1}),
 \qquad k=1,\ldots,n.
 \label{app:eq:slow-spectrum-limit}
\end{equation}
All other eigenvalues have the form \(\Gamma\nu+O(1)\), where
\(\nu<0\) is a nonzero eigenvalue of \(G_{\rm sym}\).
Together with monotonicity, this proves
\begin{equation}
 g_{\rm aut}(\Gamma)\nearrow g(\Reff)
 \qquad(\Gamma\to\infty).
 \label{app:eq:joint-gap-limit}
\end{equation}
If \(g(\Reff)>g_{\rm st}^{\rm opt}\), the strict inequality
\(g_{\rm aut}(\Gamma)>g_{\rm st}^{\rm opt}\) therefore holds
for all sufficiently large finite \(\Gamma\).

\subsection{Optimal allocation}
The spectral bound and limit in
Eqs.~\eqref{app:eq:finite-upper} and~\eqref{app:eq:joint-gap-limit}
show that, for each fixed device, the supremum of its finite-rate
joint gap equals the gap of its averaged generator.
It remains to optimize the latter over admissible mixtures.

Consider any finite mixture of profiles
\(\boldsymbol\delta^{(c)}\in\mathcal A_\Delta\), with weights
\(q_c^*\). Convexity of the exponential gives, for
\(0\leq z\leq\Delta\),
\begin{equation}
 e^{\beta z}\leq1+(e^x-1)\frac{z}{\Delta}.
 \label{app:eq:exponential-chord}
\end{equation}
Set \(r_e=\Delta^{-1}\sum_cq_c^*\delta_e^{(c)}\).
Since \(r_e\geq0\) and \(\sum_e r_e\leq1\), choose
\(\alpha_e\geq r_e\) with \(\sum_e\alpha_e=1\).
Equation~\eqref{app:eq:exponential-chord} then yields
\begin{equation}
 \overline w_e
 =w_e^0\sum_cq_c^*e^{\beta\delta_e^{(c)}}
 \leq w_e^0[1+(e^x-1)\alpha_e]
 \equiv w_e(\boldsymbol\alpha).
 \label{app:eq:extreme-domination}
\end{equation}
The conductances on the right are attainable. For each edge \(e\)
with \(\alpha_e>0\), introduce a configuration of equilibrium weight
\(\alpha_e\) whose profile is
\begin{equation}
 \delta_{e'}^{(e)}=\Delta\,\delta_{ee'},
 \label{app:eq:extreme-profile}
\end{equation}
where \(\delta_{ee'}\) is the Kronecker delta. On this positive support,
\(\widehat G=\boldsymbol\alpha\boldsymbol 1^{\mathsf T}-I\)
is irreducible and reversible with stationary distribution
\(\boldsymbol\alpha\).

Monotonicity of the gap and the explicit construction above show that the
maximum averaged gap is attained within this set. The gap is continuous
on the compact simplex of weights, so this maximum exists.
Conductance monotonicity also permits a frozen optimum to spend the
full budget; that optimum is included as a one-configuration device.
Together with the finite-rate bound and limit above, this gives, with
\(\mathcal S_E=\{\boldsymbol\alpha:\alpha_e\geq0,
\sum_e\alpha_e=1\}\),
\begin{equation}
 g_{\rm aut}^{\sup}
 =\max_{\boldsymbol\alpha\in\mathcal S_E}
 g\!\left(\{w_e^0[1+(e^x-1)\alpha_e]\}_e\right)
 \geq g_{\rm st}^{\rm opt},
 \label{app:eq:extreme-optimum}
\end{equation}
which is Eq.~\eqref{eq:general-design}. This optimization allows the auxiliary
weights and number of configurations to be chosen freely; at most one
configuration per visible edge is required. At \(\Delta=0\), both
optimal values equal the bare gap.

To obtain the semidefinite formulation, let \(u_i=\sqrt{\pi_i}\) and
\(\boldsymbol b_e=\boldsymbol e_i/\sqrt{\pi_i}
-\boldsymbol e_j/\sqrt{\pi_j}\), where \(\boldsymbol e_i\) is a
standard basis vector. The positive symmetrized generator is
\begin{equation}
 H(\boldsymbol\alpha)
 =\sum_e w_e(\boldsymbol\alpha)
 \boldsymbol b_e\boldsymbol b_e^{\mathsf T}.
 \label{app:eq:design-laplacian}
\end{equation}
Since \(H\boldsymbol u=0\) and \(\|\boldsymbol u\|=1\), requiring
every eigenvalue on \(\boldsymbol u^\perp\) to be at least \(t\)
gives
\begin{equation}
 \begin{aligned}
 &\underset{t,\,\boldsymbol\alpha}{\operatorname{maximize}}
 \quad t,\\
 &\text{subject to}\quad
 \boldsymbol\alpha\in\mathcal S_E,\qquad
 H(\boldsymbol\alpha)\succeq
 t(I-\boldsymbol u\boldsymbol u^{\mathsf T}).
 \end{aligned}
 \label{app:eq:design-sdp}
\end{equation}
Its optimal value is \(g_{\rm aut}^{\sup}\).

\subsection{Matching cost}
For each edge \(e=\{i,j\}\), Eq.~\eqref{app:eq:Reff} and
the conductance relation in Eq.~\eqref{eq:budget} give
\begin{equation}
 \overline w_e
 =\sum_cq_c^*w_e^{(c)}
 =w_e^0\sum_cq_c^*e^{\beta\delta_e^{(c)}}.
 \label{app:eq:average-conductance}
\end{equation}
A frozen profile reproducing this conductance requires the edge lowering
\(\delta_e^{\rm match}
=\frac{1}{\beta}\ln\frac{\overline w_e}{w_e^0}\).
The auxiliary distribution conditioned on an equilibrium crossing
of edge \(e\) is
\begin{equation}
 q_e^\ddagger(c)
 =\frac{q_c^*w_e^{(c)}}{\overline w_e}
 =q_c^*e^{\beta[\delta_e^{(c)}-\delta_e^{\rm match}]}.
 \label{app:eq:TS-biased}
\end{equation}
Substituting this expression into the definition of relative entropy yields
\begin{equation}
 \begin{aligned}
 \DKL(q^*\Vert q_e^\ddagger)
 &=\sum_cq_c^*\ln\frac{q_c^*}{q_e^\ddagger(c)}\\
 &=\beta\delta_e^{\rm match}
   -\beta\sum_cq_c^*\delta_e^{(c)}.
 \end{aligned}
 \label{app:eq:edge-emulation-identity}
\end{equation}
Here each conditional profile is assumed to use the full budget,
\(\sum_e\delta_e^{(c)}=\Delta\). Summing
Eq.~\eqref{app:eq:edge-emulation-identity} over the edges and using
\(\sum_cq_c^*=1\) therefore gives
\begin{equation}
 \Delta_{\rm match}
 \equiv\sum_e\delta_e^{\rm match}
 =\Delta+\kB T\sum_e\DKL(q^*\Vert q_e^\ddagger)
 \geq\Delta,
 \label{app:eq:emulation-theorem}
\end{equation}
which is Eq.~\eqref{eq:matching-cost}. Equality holds if and only if
\(q_e^\ddagger=q^*\) for every edge, equivalently if
\(\delta_e^{(c)}\) is independent of \(c\) on the support of \(q^*\).
For an unsaturated mixture, the same identity holds with the first
\(\Delta\) on the right replaced by
\(\sum_cq_c^*\sum_e\delta_e^{(c)}\leq\Delta\); exceeding the allowed
frozen budget is therefore not automatic for every admissible mixture.

\section{Finite-rate corrections and auxiliary departure}
\label{app:hierarchy}
\label{sec:hierarchy}

For this section, the device, normalized auxiliary generators, and
full-support product initial preparation are fixed. Only the common
auxiliary switching scale $\Gamma$ varies. We use the symmetrization
and slow--fast decomposition constructed in Appendix~\ref{app:optimization}.

\subsection{Rate correction}
Choose an orthonormal basis adapted to
\(\mathcal V_{\rm s}=\ker G_{\rm sym}\) and
\(\mathcal V_{\rm f}=\mathcal V_{\rm s}^{\perp}\).
The symmetrized generator in Eq.~\eqref{app:eq:symmetrization}
then has the block form
\begin{equation}
 \cL_\Gamma^{\rm sym}
 =\begin{pmatrix}
 A & B^{\mathsf T}\\
 B & C-\Gamma H
 \end{pmatrix}.
 \label{app:eq:block-generator}
\end{equation}
Here \(H=-G_{\rm sym}|_{\mathcal V_{\rm f}}>0\), and \(A\)
is orthogonally similar to
\(D_\pi^{-1/2}\Reff D_\pi^{1/2}\).
Set \(g_{\rm aut}^{\infty}=g(\Reff)\), and let \(u_2\) be a normalized
eigenvector of \(A\) with simple eigenvalue
\(\lambda_2=-g_{\rm aut}^{\infty}<0\).
For the corresponding joint eigenvalue \(\lambda_\Gamma=-r_{\rm vis}(\Gamma)\),
choose a normalized eigenvector with coordinates
\((u_\Gamma,v_\Gamma)\) and \(u_\Gamma\to u_2\).
The fast block gives
\begin{equation}
 v_\Gamma=(\Gamma H-C+\lambda_\Gamma I)^{-1}Bu_\Gamma.
 \label{app:eq:fast-component-exact}
\end{equation}
Substitution into the slow block yields
\begin{equation}
 \left[A+B^{\mathsf T}
 (\Gamma H-C+\lambda_\Gamma I)^{-1}B\right]u_\Gamma
 =\lambda_\Gamma u_\Gamma.
 \label{app:eq:Schur}
\end{equation}
For large \(\Gamma\),
\begin{equation}
 (\Gamma H-C+\lambda_\Gamma I)^{-1}
 =\Gamma^{-1}H^{-1}+O(\Gamma^{-2}).
 \label{app:eq:resolvent}
\end{equation}
Since the averaged gap is simple and all fast decay rates
diverge with $\Gamma$, this branch is the joint gap,
$r_{\rm vis}(\Gamma)=g_{\rm aut}(\Gamma)$, for sufficiently
large $\Gamma$. First-order perturbation theory applied to
Eq.~\eqref{app:eq:Schur} therefore gives
\begin{equation}
 g_{\rm aut}^{\infty}-g_{\rm aut}(\Gamma)
 =\frac{\mu_2}{\Gamma}+O(\Gamma^{-2}),
 \label{app:eq:rate-hierarchy}
\end{equation}
where \(\mu_2=(Bu_2)^{\mathsf T}H^{-1}(Bu_2)\geq0\).
This is the first relation in Eq.~\eqref{eq:hierarchy}.
The coefficient may vanish; the branch is identified with the joint gap
only in the large-$\Gamma$ regime.

\subsection{Uniform peak-departure bound}
Consider a fixed finite device and a fixed product initial state
\(P(0)=p(0)\otimes q^*\), with \(p_i(0)>0\) for every \(i\).
Using the projection \(\Q=I-\E\M\) defined in Appendix~\ref{app:averaging}, write
\(Y=\Q P=P-\E p\).
Then \(Y(0)=0\). Since \(\Q G=G\) and \(G\E=0\),
applying \(\Q\) to the joint equation gives
\(\dot Y=\Gamma GY+\Q\LS P\).
Irreducibility and reversibility of the conditional auxiliary generators
give
\begin{equation}
 \|e^{\Gamma Gt}\Q\|_1\leq C_Ge^{-\Gamma h_*t},
 \label{app:eq:auxiliary-semigroup-bound}
\end{equation}
where \(h_*>0\) is their smallest spectral gap and \(C_G\) is
independent of time and \(\Gamma\).
The integral solution is
\begin{equation}
 Y(t)=\int_0^t e^{\Gamma G(t-s)}\Q\LS P(s)\,ds.
 \label{app:eq:Y-integral}
\end{equation}
Let \(K=\|\LS\|_1=\max_{j,c}\sum_i|R_{ij}^{(c)}|\), which is
independent of \(\Gamma\). Since \(\|P(s)\|_1=1\),
Eq.~\eqref{app:eq:auxiliary-semigroup-bound} yields
\begin{equation}
 \sup_{t\geq0}\|Y(t)\|_1
 \leq\frac{C_GK}{\Gamma h_*}.
 \label{app:eq:Y-scaling}
\end{equation}
Set \(m_0=\min_i[p_i(0)/\pi_i]>0\).
The inequality \(P(0)\geq m_0\Pi\) and positivity of the Markov
semigroup imply \(P(t)\geq m_0\Pi\), and hence
\(p_i(t)q_c^*\geq m_0\Pi_{ic}\), uniformly in time and \(\Gamma\).
Using \(\ln z\leq z-1\) gives
\begin{equation}
 \Dt(t)
 \leq\sum_{i,c}\frac{Y_{ic}(t)^2}{p_i(t)q_c^*}
 \leq\frac{\|Y(t)\|_1^2}{m_0\min_{i,c}\Pi_{ic}}.
 \label{app:eq:departure-quadratic-bound}
\end{equation}
Combining Eqs.~\eqref{app:eq:Y-scaling} and
\eqref{app:eq:departure-quadratic-bound} proves
\begin{equation}
 \widetilde D_{\max}(\Gamma)
 \equiv\max_{t\geq0}\Dt_\Gamma(t)
 =O(\Gamma^{-2}).
 \label{app:eq:peak-memory}
\end{equation}
Together with Eq.~\eqref{app:eq:rate-hierarchy}, this reproduces
Eq.~\eqref{eq:hierarchy}.
The constant in the departure bound depends on the fixed device and
initial preparation, but not on time or \(\Gamma\).
If \(g_{\rm aut}^{\infty}>g_{\rm st}^{\rm opt}\), convergence of
the joint gap and this uniform bound imply that, for every
\(\varepsilon>0\), all sufficiently large finite \(\Gamma\)
satisfy both \(g_{\rm aut}(\Gamma)>g_{\rm st}^{\rm opt}\) and
\(\widetilde D_{\max}(\Gamma)<\varepsilon\).

\subsection{Information and free-energy interpretation}\label{sec:entropy}
Let \(p_i=\sum_cP_{ic}\), \(q_c=\sum_iP_{ic}\), and
\((\E p)_{ic}=p_iq_c^*\). Using natural logarithms, define
\begin{equation}
 \Dt\equiv\DKL(P\Vert\E p)
 =\sum_{i,c}P_{ic}\ln\frac{P_{ic}}{p_iq_c^*}.
 \label{app:eq:memory-def}
\end{equation}
Splitting the logarithm through \(p_iq_c\) gives
\begin{equation}
 \begin{aligned}
 \Dt
 &=\sum_{i,c}P_{ic}\ln\frac{P_{ic}}{p_iq_c}
   +\sum_cq_c\ln\frac{q_c}{q_c^*}\\
 &=I(S{:}C)+\DKL(q\Vert q^*),
 \end{aligned}
 \label{app:eq:memory-chain}
\end{equation}
so $\Dt$ measures both instantaneous mutual information and displacement
of the auxiliary marginal from equilibrium. For state-independent
auxiliary switching, $\dot q=\Gamma\widehat Gq$; consequently
$q(0)=q^*$ implies $q(t)=q^*$ and $\Dt(t)=I(S{:}C)$.

\subsection{Entropy and free-energy balances}
For $D_S=\DKL(p\Vert\pi)$, the chain rule gives
\begin{equation}
 \DKL(P\Vert\Pi)=D_S+\Dt.
 \label{app:eq:joint-chain}
\end{equation}
With the additive basin energies in Eq.~\eqref{app:eq:additive-minima},
the nonequilibrium free energy~\cite{Ge2010}
$\mathcal F[X]=\sum_{i,c}X_{ic}[F_{ic}+\kB T\ln X_{ic}]$ obeys
$\mathcal F[X]-\mathcal F[\Pi]=\kB T\DKL(X\Vert\Pi)$.
Since $\DKL(\E p\Vert\Pi)=D_S$, this yields
$\mathcal F[P]-\mathcal F[\E p]=\kB T\Dt$ at fixed $p$.
For reversible relaxation~\cite{Sekimoto2010},
$\Sigma_{\rm relax}(t)/\kB=
\DKL[P(0)\Vert\Pi]-\DKL[P(t)\Vert\Pi]$.
A product preparation has $\Dt(0)=0$, hence
\begin{equation}
 D_S(0)-D_S(t)=\Sigma_{\rm relax}(t)/\kB+\Dt(t).
 \label{app:eq:thermo-product}
\end{equation}
At complete relaxation both terms on the right of
Eq.~\eqref{app:eq:joint-chain} vanish, giving
$\Sigma_{\rm relax}(\infty)=\kB D_S(0)$ for every $\Gamma>0$.

\section{Three-state witness and numerical validation}
\label{app:three-state}
\label{sec:finite}

We derive the three-state gap ratio in Eq.~\eqref{eq:minimal-witness} and the
finite-rate threshold, then give the numerical checks for
Figs.~\ref{fig:resource} and~\ref{fig:finite}.

\subsection{Optimal gaps and budget constraints}
For the path \(0\leftrightarrow1\leftrightarrow2\), with
\(\pi_i=1/3\) and edge conductances \((u,v)\), the column generator is
\begin{equation}
 R(u,v)=3
 \begin{pmatrix}
 -u & u & 0\\
 u & -(u+v) & v\\
 0 & v & -v
 \end{pmatrix},
 \label{app:eq:path-generator}
\end{equation}
with gap
\begin{equation}
 g(u,v)=3\left[u+v-\sqrt{u^2-uv+v^2}\right].
 \label{app:eq:path-gap}
\end{equation}
Set \(x=\beta\Delta>0\), \(a=w_0e^x\), and \(b=w_0\).
Monotonicity permits the frozen optimum to use the full budget, so
\(uv=ab\). At fixed product, the gap decreases with \(s=u+v\),
since \(\partial_s g=3[1-s/\sqrt{s^2-3ab}]<0\).
Its maximum is therefore attained at \(u=v=\sqrt{ab}\), giving
\(g_{\rm st}^{\rm opt}=3\sqrt{ab}\) [Fig.~\ref{fig:resource}(a)].
By the optimal-design result in Appendix~\ref{app:optimization}, autonomous optimization reduces
to mixing \((a,b)\) and \((b,a)\) [Fig.~\ref{fig:resource}(b)]. These mixtures have
\(\overline u+\overline v=a+b\); Eq.~\eqref{app:eq:path-gap} is
maximized at fixed sum by equal conductances. Equal mixing thus gives
\(g_{\rm aut}^{\infty}=g_{\rm aut}^{\sup}=3(a+b)/2\), and
\begin{equation}
 \frac{g_{\rm aut}^{\sup}}{g_{\rm st}^{\rm opt}}
 =\frac{a+b}{2\sqrt{ab}}
 =\cosh\!\left(\frac{x}{2}\right)>1.
 \label{app:eq:path-gap-ratio}
\end{equation}
This is the gain shown in Fig.~\ref{fig:resource}(c).

Reproducing these averaged conductances with a frozen profile requires
\begin{equation}
 \Delta_{\rm match}
 =\frac{2}{\beta}\ln\frac{1+e^x}{2}
 =\Delta+\frac{2}{\beta}\ln\cosh\!\left(\frac{x}{2}\right),
 \label{app:eq:path-matching-budget}
\end{equation}
consistent with Eq.~\eqref{eq:matching-cost}. If the frozen class is convex in
conductance space at fixed \(\pi\) and contains every conditional
profile, it contains their average. Its gap bounds the joint gap by
Eq.~\eqref{app:eq:finite-upper}, so switching cannot exceed the best
frozen gap. Here \(\Delta_{\rm match}>\Delta\) puts the averaged
conductances outside the allowed frozen set.

For an additional edgewise lowering cap \(\Delta/2\leq d\leq\Delta\),
the balanced frozen optimum remains feasible. Equally mixing the lowering
profiles \((d,\Delta-d)\) and \((\Delta-d,d)\) gives
\begin{equation}
 \frac{g_{\rm aut}^{\infty,(d)}}{g_{\rm st}^{\rm opt}}
 =\cosh[\beta(d-\Delta/2)].
 \label{app:eq:capped-gap-ratio}
\end{equation}
The gain persists for \(d>\Delta/2\) and vanishes at \(d=\Delta/2\).

\subsection{Exact spectrum and switching threshold}
For the uncapped profiles \((a,b)\) and \((b,a)\), take equal auxiliary
weights and transition rates \(\Gamma/2\) in both directions.
In the ordering \((0A,1A,2A,0B,1B,2B)\), the joint generator is
\begin{equation}
 \cL_\Gamma
 =\operatorname{diag}[R(a,b),R(b,a)]
 +\frac{\Gamma}{2}
 \begin{pmatrix}-1&1\\1&-1\end{pmatrix}\otimes I_3.
 \label{app:eq:three-state-joint}
\end{equation}
With \(s=a+b\) and \(r=\sqrt{ab}\), its characteristic polynomial is
\begin{equation}
 \begin{aligned}
 \det(\lambda I-\cL_\Gamma)
 &=\frac{\lambda(\lambda+\Gamma)}{4}Q_1(\lambda)Q_2(\lambda),\\
 Q_1(\lambda)
 &=2\lambda^2+(12s+2\Gamma)\lambda+54r^2+3\Gamma s,\\
 Q_2(\lambda)&=Q_1(\lambda)+6\Gamma s.
 \end{aligned}
 \label{app:eq:joint-characteristic}
\end{equation}
The smaller decay rate from \(Q_1\) is
\begin{equation}
 r_{\rm vis}(\Gamma)
 =\frac{\Gamma+6s
 -\sqrt{\Gamma^2+6\Gamma s+36(s^2-3r^2)}}{2}.
 \label{app:eq:exact-visible-rate}
\end{equation}
The corresponding square-root argument for \(Q_2\) is
\((\Gamma-3s)^2+27(a-b)^2\), smaller by \(12\Gamma s\), so its
smaller decay rate is no less than \(r_{\rm vis}\). The remaining
quadratic roots give larger rates. Including the auxiliary eigenvalue
\(-\Gamma\) therefore yields the full joint gap
\begin{equation}
 g_{\rm aut}(\Gamma)=\min\{\Gamma,r_{\rm vis}(\Gamma)\}.
 \label{app:eq:full-joint-gap}
\end{equation}

To compare with \(g_{\rm st}^{\rm opt}=3r\), note that
\(\Gamma+6s-6r>0\). Squaring the inequality
\(r_{\rm vis}(\Gamma)>3r\) then reduces it to
\[
 \begin{aligned}
 &[\Gamma+6s-6r]^2
 -[\Gamma^2+6\Gamma s+36(s^2-3r^2)]\\
 &\qquad=6(s-2r)(\Gamma-12r)>0.
 \end{aligned}
\]
Since \(x>0\) implies \(s>2r\), this holds exactly for
\(\Gamma>12r\), which also ensures \(\Gamma>3r\). Hence
\begin{equation}
 g_{\rm aut}(\Gamma)>g_{\rm st}^{\rm opt}
 \quad\Longleftrightarrow\quad
 \Gamma>\Gamma_c=12\sqrt{ab}=4g_{\rm st}^{\rm opt}.
 \label{app:eq:exact-threshold}
\end{equation}
This gives the threshold in Fig.~\ref{fig:finite}(a).
At large \(\Gamma\), the joint gap equals
\(r_{\rm vis}(\Gamma)\), and expansion gives
\[
 g_{\rm aut}^{\infty}-g_{\rm aut}(\Gamma)
 =\frac{27(a-b)^2}{4\Gamma}+O(\Gamma^{-2}).
\]
Thus this model realizes Eq.~\eqref{eq:hierarchy} with \(\mu_2=27(a-b)^2/4\).

\subsection{Exact dynamics and trajectory validation}\label{sec:numerics}
The three-state plots in Fig.~\ref{fig:finite} use \(x=1\), \(w_0=1\),
\(g_0=3w_0\), \(q^*=(1/2,1/2)\), and the full-support preparation
\(p(0)=(0.80,0.15,0.05)^{\mathsf T}\), with
\(P(0)=p(0)\otimes q^*\). Times and rates are reported as
\(g_0t\) and \(\Gamma/g_0\).
Diagonalizing Eq.~\eqref{app:eq:three-state-joint} gives the joint gap
[Fig.~\ref{fig:finite}(a)] and rate deficit [Fig.~\ref{fig:finite}(c)]. Spectral propagation gives
\(\DKL[p(t)\Vert\pi]\) [Fig.~\ref{fig:finite}(b)] and the peak auxiliary departure
[Figs.~\ref{fig:finite}(a) and~\ref{fig:finite}(c)].
The visible branch agrees with Eq.~\eqref{app:eq:exact-visible-rate}
to machine precision.
The time grid resolves both \(\Gamma^{-1}\) and \(g_0^{-1}\);
doubling its logarithmic and linear densities changes
\(\Dt_{\max}\) by less than the plotted line width.

\begin{figure*}[!htbp]
\centering
\begin{minipage}[t]{0.48\textwidth}
 \centering
 \includegraphics[width=\linewidth]{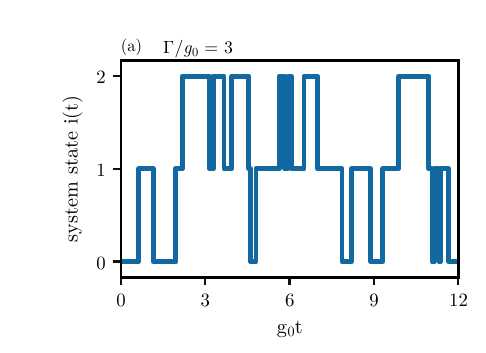}
\end{minipage}\hspace{0.012\textwidth}
\begin{minipage}[t]{0.48\textwidth}
 \centering
 \includegraphics[width=\linewidth]{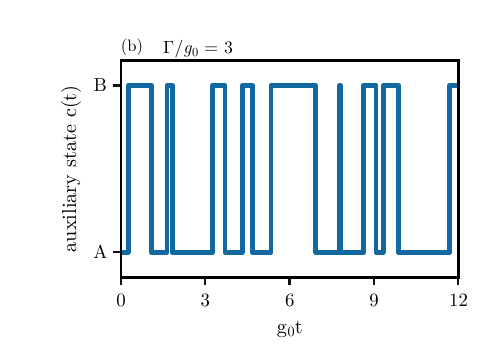}
\end{minipage}

\vspace{0.15em}
\begin{minipage}[t]{0.48\textwidth}
 \centering
 \includegraphics[width=\linewidth]{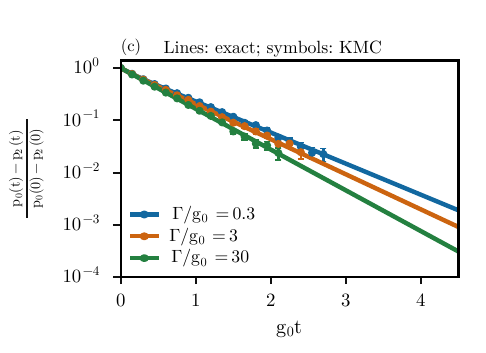}
\end{minipage}\hspace{0.012\textwidth}
\begin{minipage}[t]{0.48\textwidth}
 \centering
 \includegraphics[width=\linewidth]{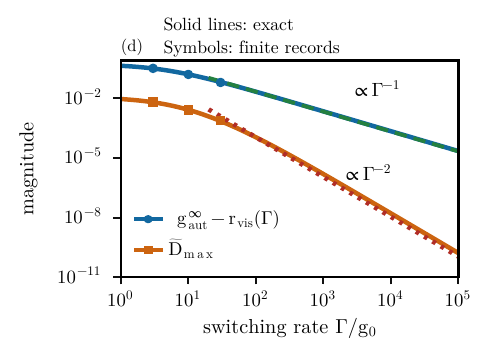}
\end{minipage}
\caption{\textbf{Trajectory-level kinetic-Monte-Carlo simulation.}
(a),(b) System and auxiliary components of one illustrative trajectory at
\(\Gamma/g_0=3\). (c) Exact normalized imbalance (lines) and Monte Carlo
estimates (symbols) for three switching rates. Error bars are one standard error;
points are shown where the signal exceeds three standard errors. (d) Exact rate deficit
\([g_{\rm aut}^{\infty}-r_{\rm vis}(\Gamma)]/g_0\) (with \(g_0=1\)) and peak
\(\widetilde D\) (stored free energy in units of \(k_{\rm B}T\)), together
with finite-record reconstructions. Peak-departure error bars are bootstrap
standard deviations. The guides show the predicted
\(\Gamma^{-1}\) and \(\Gamma^{-2}\) powers.}
\label{app:fig:kmc-validation}
\end{figure*}

Independent Gillespie simulations~\cite{Gillespie1976} sample the same
time-homogeneous generator without an imposed switching schedule.
In units \(g_0=1\), the lowered and bare visible transition rates are
\(e\) and \(1\), and auxiliary transitions have rate \(\Gamma/2\).
Figure~\ref{app:fig:kmc-validation}(a) follows the visible state and
panel (b) the auxiliary configuration in one trajectory at
\(\Gamma/g_0=3\) over \(g_0T=12\), below the optimal-frozen crossover.
Panel (c) compares the exact imbalance \(m(t)=p_0(t)-p_2(t)\),
which overlaps the visible slow mode, with estimates from
\(3\times10^4\) independent product-initialized trajectories at each
of \(\Gamma/g_0=0.3,3,30\), sampled at
\(g_0t=0,0.075,\ldots,4.5\). Error bars are one standard error.

Figure~\ref{app:fig:kmc-validation}(d) tests finite-record reconstruction
at \(\Gamma/g_0=3,10,30\). For the rate, we use 32 independent
equilibrium-initialized trajectories of duration \(g_0T=2000\) at each
switching rate. On the symmetric six-state topology, jump counts
\(N_{ij}\) from \(j\) to \(i\) and residence times \(T_i\) give the
reversible estimator
\begin{equation}
 \widehat r_{ij}=\widehat r_{ji}
 =\frac{N_{ij}+N_{ji}}{T_i+T_j}.
 \label{app:eq:reversible-rate-estimator}
\end{equation}
We report the mean and standard error of the reconstructed full gaps
across records. At these switching rates,
the exact joint gap equals \(r_{\rm vis}(\Gamma)\).

For the peak departure, \(10^5\) independent trajectories start from
the specified non-equilibrium product preparation at each switching rate.
Joint states are recorded at 240 times resolving early auxiliary and
later visible relaxation. We divide trajectories into 100 groups of
1000 and draw 2000 bootstrap samples of 100 groups with replacement,
retaining all observation times within each group to preserve temporal
correlations. For each sample, we recompute \(\Dt(t)\) from observed
frequencies and take its peak; the standard deviation of these peaks
sets the error bar. Finite sampling and maximization can bias the peak
estimates upward. The asymptotic laws in Eq.~\eqref{eq:hierarchy} are proved in Appendix~\ref{app:hierarchy}.

\section{Collective magnetization dynamics and frozen bounds}
\label{app:collective}
\label{sec:collective}\label{sec:frozen-bounds}

This appendix gives the global optimization behind Fig.~\ref{fig:validation}(a), the
finite-rate test of the same optimal device in Fig.~\ref{fig:validation}(b), and a
simpler two-configuration reference.

\subsection{Magnetization states and bare dynamics}
For \(N\) spins \(s_k=\pm1\), the total magnetization
\(m=\sum_{k=1}^N s_k\) takes the values \(-N,-N+2,\ldots,N\).
The energy, sector multiplicity, and equilibrium distribution are
\begin{align}
 E_N(m)&=-\frac{J}{2N}m^2-hm,
 \label{app:eq:CWenergy}\\
 \Omega_N(m)&=\binom{N}{(N+m)/2},
 \label{app:eq:CWmult}\\
 \pi_N(m)&=Z_N^{-1}\Omega_N(m)e^{-\beta E_N(m)}.
 \label{app:eq:CWpi}
\end{align}
With \(\Delta E_m=E_N(m+2)-E_N(m)\), summing symmetric exponential
single-spin-flip rates over the available spins gives the exactly
lumped magnetization generator,
\begin{align}
 R^{(0)}_{m+2,m}
 &=\frac{N-m}{2}\gamma e^{-\beta\Delta E_m/2},
 \label{app:eq:CWup}\\
 R^{(0)}_{m,m+2}
 &=\frac{N+m+2}{2}\gamma e^{\beta\Delta E_m/2},
 \label{app:eq:CWdown}
\end{align}
where \(\gamma\) is the attempt rate. All edge indices below run over
\(m=-N,-N+2,\ldots,N-2\); the diagonal entries make each column sum
to zero. The identity
\(\Omega_N(m+2)/\Omega_N(m)=(N-m)/(N+m+2)\) gives
\(\pi_N(m)R^{(0)}_{m+2,m}=\pi_N(m+2)R^{(0)}_{m,m+2}\).
The chain is therefore reversible with stationary distribution
\(\pi_N\). Its gap describes magnetization relaxation and need not
equal the gap of the full spin-configuration dynamics. Each
magnetization edge has one fixed aggregated bare conductance: the
spin multiplicities in Eqs.~\eqref{app:eq:CWup} and~\eqref{app:eq:CWdown}
are not additional adjustable transition channels.

\subsection{Budget and a two-configuration reference}
Two complementary barrier-lowering profiles specify the budget
and define a simple reference device:
\begin{align}
 \Delta_m^{(A)}
 &=\Delta_0\exp\!\left[-\frac{(m+1+m_c)^2}{2\sigma^2N^2}\right],
 \nonumber\\
 \Delta_m^{(B)}
 &=\Delta_0\exp\!\left[-\frac{(m+1-m_c)^2}{2\sigma^2N^2}\right],
 \qquad m_c=0.25N.
 \label{app:eq:CWdrop}
\end{align}
For \(c=A,B\), both rates on edge \(m\leftrightarrow m+2\) are
multiplied by \(e^{\beta\Delta_m^{(c)}}\), producing \(R^{(c)}\)
with the same equilibrium distribution \(\pi_N\). Reflection
\(m\mapsto-m-2\) interchanges the profiles, so each configuration
uses the same total budget,
\(B_N=\sum_m\Delta_m^{(A)}=\sum_m\Delta_m^{(B)}\), denoted
by \(\Delta\) above.
A frozen profile \(\boldsymbol\delta\) multiplies both bare rates on
edge \(m\) by \(e^{\beta\delta_m}\). Its generator is
\(R(\boldsymbol\delta)\), and the comparison includes every allocation
in the simplex
\begin{equation}
 \begin{gathered}
 \mathcal A_N=\left\{\boldsymbol\delta:\delta_m\geq0,
 \quad\sum_m\delta_m\leq B_N\right\},\\
 g_{\rm st}^{\rm opt}=\max_{\boldsymbol\delta\in\mathcal A_N}
 g[R(\boldsymbol\delta)].
 \end{gathered}
 \label{app:eq:CWstaticclass}
\end{equation}
The frozen class thus allows independent allocation to all \(N\) edges.
We use
\(\beta=1,\qquad J=1.2,\qquad h=0,\qquad \gamma=1,\qquad\Delta_0=2.5,\qquad\sigma=0.15\).
The profile width is proportional to \(N\), giving \(B_N=O(N)\).
Budgets are matched at each size, with no individual-edge lowering cap.
The parameter \(\Delta_0\) limits only the two reference profiles.

\subsection{Globally optimal fast-switching design}
The bare edge conductances are
\(w_m^0=\pi_N(m)R^{(0)}_{m+2,m}\), and \(g_0=g(R^{(0)})\).
At each fixed \(N\), apply Eq.~\eqref{app:eq:extreme-optimum} with
\(\Delta=B_N\). Relabel the magnetization states by \(i=0,\ldots,N\)
and their edges by \(i=0,\ldots,N-1\), with
\(\pi_i=\pi_N(-N+2i)\). Set \(K_N=e^{\beta B_N}-1\). The
optimization is over the averaged conductances
\begin{equation}
 \overline w_i=w_i^0(1+K_N\alpha_i),\qquad
 \alpha_i\geq0,\qquad\sum_{i=0}^{N-1}\alpha_i=1.
 \label{app:eq:CWoptimal-class}
\end{equation}
Configuration \(i\) places the full lowering \(B_N\) on edge \(i\),
with auxiliary equilibrium weight \(\alpha_i\).

The optimum has an explicit form when the weights below are nonnegative.
Define
\begin{equation}
 \begin{gathered}
 F_0=0,\qquad F_{i+1}-F_i=(w_i^0)^{-1/2},\\
 f_i=F_i-\sum_{j=0}^{N}\pi_jF_j,\qquad
 V_N=\sum_{i=0}^{N}\pi_if_i^2.
 \end{gathered}
 \label{app:eq:CWoptimal-trial}
\end{equation}
For every feasible \(\boldsymbol\alpha\), the Rayleigh principle gives
the same upper bound,
\begin{equation}
 g(\overline{\boldsymbol w})
 \leq\frac{\sum_{i=0}^{N-1}\overline w_i(f_{i+1}-f_i)^2}{V_N}
 =\frac{N+K_N}{V_N}\equiv g_*.
 \label{app:eq:CWoptimal-upper}
\end{equation}
Introduce
\begin{equation}
 \zeta_i=-\frac{\sum_{j=0}^{i}\pi_jf_j}{\sqrt{w_i^0}},\qquad
 w_i^*=g_*w_i^0\zeta_i,\qquad
 \alpha_i^*=\frac{g_*\zeta_i-1}{K_N}.
 \label{app:eq:CWoptimal-weights}
\end{equation}
The centered function \(f\) is strictly increasing, so each proper
prefix sum \(\sum_{j=0}^{i}\pi_jf_j\) is negative and \(\zeta_i>0\).
Discrete summation by parts gives \(\sum_i\zeta_i=V_N\), and hence
\(\sum_i\alpha_i^*=1\). If \(\alpha_i^*\geq0\) for every edge,
these weights are feasible. Moreover,
\(w_i^*(f_{i+1}-f_i)=-g_*\sum_{j=0}^{i}\pi_jf_j\), so the
corresponding backward generator obeys
\(R(\boldsymbol w^*)^{\mathsf T}f=-g_*f\).
Taking successive differences of the backward eigenvalue equation
gives an irreducible tridiagonal operator with negative off-diagonal
entries and the nonzero spectrum of \(-R^{\mathsf T}\). Its positive
eigenvector \(f_{i+1}-f_i\) identifies \(g_*\) as the gap by the
Perron--Frobenius theorem.
Thus the bound in Eq.~\eqref{app:eq:CWoptimal-upper} is attained,
\begin{equation}
 g_{\rm aut}^{\sup}=\frac{N+e^{\beta B_N}-1}{V_N},
 \qquad \text{provided }\min_i\alpha_i^*\geq0.
 \label{app:eq:CWglobal-optimum}
\end{equation}
The positive weights in Table~\ref{app:tab:CWoptimal} verify the optimum
for all four sizes using \(N\) auxiliary configurations.
The supremum over finite switching rates equals this limiting gap;
the construction does not assert that a finite rate attains it.
Independent diagonalization of the symmetric tridiagonal generator
agrees with Eq.~\eqref{app:eq:CWglobal-optimum} to relative error
below \(4\times10^{-14}\) for these sizes.

\begin{table*}[t]
\caption{Global autonomous optimum for the Curie--Weiss magnetization
chain. The positive minimum weight verifies applicability of
Eq.~\eqref{app:eq:CWglobal-optimum}. The final column uses the frozen
upper bound in Eq.~\eqref{app:eq:CWcertificate}, and therefore gives
a lower bound on the gain over the true frozen optimum, rounded down.}
\label{app:tab:CWoptimal}
\centering
\begin{ruledtabular}
\begin{tabular}{c c c c c}
\(N\)&\(\beta B_N\)&\(g_{\rm aut}^{\sup}/g_0\)&
\(\min_i\alpha_i^*\)&\(g_{\rm aut}^{\sup}/g_{\rm st}^{\rm ub}\)\\
\hline
10&4.70&13.3&0.0431&7.03\\
20&9.40&747&0.0108&359\\
30&14.1&\(5.86\times10^4\)&0.00349&\(2.64\times10^4\)\\
40&18.8&\(5.11\times10^6\)&0.00132&\(2.19\times10^6\)\\
\end{tabular}
\end{ruledtabular}
\end{table*}

The growth reflects \(B_N=O(N)\) and the uncapped single-edge
activation factor \(e^{\beta B_N}\), not a fixed-total-budget
thermodynamic limit. Within each optimization, \(\pi_N\),
\(R^{(0)}\), and \(B_N\) are fixed.

\subsection{Global frozen-profile bound on a tree}
With \(\boldsymbol b_e\) and \(\boldsymbol u\) as defined in Appendix~\ref{app:optimization},
let \(B\) have rows \(\boldsymbol b_e^{\mathsf T}\), and let \(U\)
have orthonormal columns spanning \(\boldsymbol u^\perp\).
For a tree with \(n\) vertices, \(A=BU\) is a square invertible
matrix. The positive generator restricted to this subspace is
\begin{equation}
 \widehat H(\boldsymbol\delta)
 =A^{\mathsf T}\operatorname{diag}
 \bigl(w_e^0e^{\beta\delta_e}\bigr)A.
 \label{app:eq:tree-restricted-laplacian}
\end{equation}
Writing \(\boldsymbol c_e=A^{-1}\boldsymbol e_e\) gives
\begin{equation}
 \widehat H(\boldsymbol\delta)^{-1}
 =\sum_e\frac{e^{-\beta\delta_e}}{w_e^0}
 \boldsymbol c_e\boldsymbol c_e^{\mathsf T}.
 \label{app:eq:tree-inverse}
\end{equation}
Since \(g=\lambda_{\min}(\widehat H)\),
\begin{equation}
 -\ln g(\boldsymbol\delta)
 =\max_{\|\boldsymbol z\|=1}
 \ln\!\left[
 \sum_e\frac{(\boldsymbol z^{\mathsf T}\boldsymbol c_e)^2}{w_e^0}
 e^{-\beta\delta_e}\right].
 \label{app:eq:tree-log-concavity}
\end{equation}
The log-sum-exp is convex in \(\boldsymbol\delta\) for each
\(\boldsymbol z\), as is its pointwise maximum. Thus \(\ln g\)
is concave on \(\mathcal A_\Delta\).

Let \(\widehat{\boldsymbol\delta}\) be a feasible candidate with a
simple gap. For its slow eigenfunction \(\phi\), normalized by
\(\sum_i\pi_i\phi_i=0\) and \(\sum_i\pi_i\phi_i^2=1\),
differentiating Eq.~\eqref{app:eq:gap-variational} gives
\begin{equation}
 h_e\equiv
 \left.\partial_{\delta_e}\ln g\right|_{\widehat{\boldsymbol\delta}}
 =\frac{\beta w_e(\widehat{\boldsymbol\delta})}
 {g(\widehat{\boldsymbol\delta})}(\phi_i-\phi_j)^2\geq0.
 \label{app:eq:tree-log-gradient}
\end{equation}
Concavity then implies, for every feasible \(\boldsymbol\delta\),
\begin{equation}
 \ln g(\boldsymbol\delta)
 \leq\ln g(\widehat{\boldsymbol\delta})
 +\boldsymbol h\cdot
 (\boldsymbol\delta-\widehat{\boldsymbol\delta}).
 \label{app:eq:tree-supporting-bound}
\end{equation}
Maximizing the linear term over \(\mathcal A_\Delta\) yields
\begin{equation}
 g(\widehat{\boldsymbol\delta})
 \leq g_{\rm st}^{\rm opt}
 \leq g(\widehat{\boldsymbol\delta})e^\epsilon,
 \label{app:eq:tree-frozen-bound}
\end{equation}
where
\begin{equation}
 \epsilon=\Delta\max_e h_e
 -\widehat{\boldsymbol\delta}\cdot\boldsymbol h\geq0.
 \label{app:eq:tree-bound-residual}
\end{equation}

\subsection{Numerical frozen comparison}
The magnetization graph is a path, so
Eqs.~\eqref{app:eq:tree-log-gradient}--\eqref{app:eq:tree-bound-residual}
apply with \(\Delta=B_N\). For a feasible numerical candidate
\(\widehat{\boldsymbol\delta}\), set
\(g_{\rm st}^{\rm num}=g[R(\widehat{\boldsymbol\delta})]\).
The bound becomes
\begin{align}
 g_{\rm st}^{\rm num}\leq g_{\rm st}^{\rm opt}
 &\leq g_{\rm st}^{\rm ub}
 \equiv g_{\rm st}^{\rm num}e^{\epsilon_N},
 \label{app:eq:CWcertificate}\\
 \epsilon_N&=B_N\max_m h_m
 -\sum_m\widehat\delta_m h_m\geq0.
 \label{app:eq:CWcertificate-residual}
\end{align}
We obtain the candidate with sequential least-squares programming
(SLSQP), using the analytic gap gradient and
\(\sum_m\delta_m=B_N\), allowed by monotonicity.
Eight distinct initial profiles give final gaps agreeing within
\(10^{-9}\); the largest \(\epsilon_N\) is \(1.95\times10^{-7}\).
In Fig.~\ref{fig:validation}(a), frozen markers show \(g_{\rm st}^{\rm num}\), with
error bars extending to \(g_{\rm st}^{\rm ub}\). Their relative widths
\(e^{\epsilon_N}-1<2\times10^{-7}\) are unresolved at the plotting scale.
These endpoints evaluate the analytic bound in floating-point
arithmetic; agreement among runs is only a numerical cross-check.
The autonomous values exceed the frozen upper endpoints at every plotted
size (Table~\ref{app:tab:CWoptimal}), establishing
\(g_{\rm aut}^{\sup}>g_{\rm st}^{\rm opt}\).

\subsection{Finite switching in the globally optimal device}
For Fig.~\ref{fig:validation}(b), fix the optimal single-edge profiles and weights
\(\alpha_c^*\) at \(N=M=40\), so
\(g_{\rm aut}^{\infty}=g_{\rm aut}^{\sup}\), and choose
\begin{equation}
 \widehat G_{cc'}=\alpha_c^*-\delta_{cc'},\qquad
 \Pi_{ic}=\pi_i\alpha_c^*,
 \label{app:eq:CWoptimal-switching}
\end{equation}
in the column convention. The rate from \(c'\) to \(c\ne c'\) is
\(\Gamma\alpha_c^*\); the normalized auxiliary gap is one.
Only \(\Gamma\) varies. Let \(\varphi_i=f_i/\sqrt{V_N}\), with
\(f\) and \(V_N\) defined in Eq.~\eqref{app:eq:CWoptimal-trial},
and set \(a_0=0.20/\max_i|\varphi_i|\). The fixed preparation is
\begin{equation}
 p_i(0)=\pi_i(1+a_0\varphi_i),\qquad
 P_{ic}(0)=p_i(0)\alpha_c^*.
 \label{app:eq:CWoptimal-initial}
\end{equation}
It has full support because \(|a_0\varphi_i|\leq0.20\).
State-independent switching preserves \(q(t)=\boldsymbol\alpha^*\),
so \(\Dt(t)=I(S{:}C)\).

For the \(41\times40=1640\)-state joint generator, we compute the
gap from the exact slow--fast Schur complement and check it against
the full symmetric spectrum. In units of
\(g_{\rm aut}^{\infty}\), the positive fast block is
\(C_++\rho I\), where \(\rho=\Gamma/g_{\rm aut}^{\infty}\)
and \(C_+\succeq0\) is the compressed visible relaxation operator.
The joint gap lies below one and the fast block's eigenvalues are at least
\(\rho\). Schur-complement inertia thus identifies the second root
as the full joint gap over \(10^4\leq\rho\leq10^8\), without a
fast-switching approximation.

Spectral propagation at each of the 21 switching rates resolves the
initial auxiliary layer and visible relaxation up to
\(t=40/g_{\rm aut}^{\infty}\); local refinement determines the peak
of \(\Dt(t)\).
The bound
\begin{equation}
 \Dt(t)\leq\DKL[P(t)\Vert\Pi]
 \leq\chi^2[P(0)\Vert\Pi]e^{-2g_{\rm aut}(\Gamma)t}
 \label{app:eq:CWoptimal-tail}
\end{equation}
excludes a larger peak after the final sampled time; here
\(\chi^2(P\Vert\Pi)=\sum_{i,c}(P_{ic}-\Pi_{ic})^2/\Pi_{ic}\).
The finite-time maximization is numerical, not an interval certificate.
Writing \(z_{ic}=Y_{ic}/(p_i\alpha_c^*)\), we evaluate the departure
as \(\sum_{i,c}p_i\alpha_c^*[(1+z_{ic})\ln(1+z_{ic})-z_{ic}]\),
using a series for small \(z_{ic}\) to avoid cancellation.
Independent matrix-exponential evaluations at
\(\rho=10^4,10^6,10^8\) agree with the peak departures to relative
error below \(6.1\times10^{-10}\).

The rate coefficient from Appendix~\ref{app:hierarchy} gives
\[
 \mu_2/(g_{\rm aut}^{\infty})^2\simeq1.47\times10^4.
\]
For this initial slow-mode perturbation, define
\(h_i^{(c)}=(R^{(c)\mathsf T}\varphi)_i/g_{\rm aut}^{\infty}
+\varphi_i\). The auxiliary-equilibrated outer solution has
\(p_i(t)=\pi_i[1+a_0\varphi_i e^{-g_{\rm aut}^{\infty}t}]\)
to leading order. Each contribution to the quadratic departure is
proportional to \(z^2/(1+a_0\varphi_i z)\), which increases with
\(z=e^{-g_{\rm aut}^{\infty}t}\in[0,1]\).
In the initial layer, unit auxiliary relaxation multiplies the
leading deviation \(Y\) by \(1-e^{-\Gamma t}\), so the departure
approaches the outer coefficient from below.
Consequently,
\begin{equation}
 \lim_{\rho\to\infty}\rho^2\Dt_{\max}
 =\frac{a_0^2}{2}\sum_{i,c}
 \frac{\pi_i\alpha_c^*[h_i^{(c)}]^2}{1+a_0\varphi_i}
 \simeq83.3.
 \label{app:eq:CWoptimal-peak-coefficient}
\end{equation}
Logarithmic fits over \(10^7\leq\rho\leq10^8\) give slopes
\(-0.999\) for the relative gap deficit and \(-1.999\) for
\(\Dt_{\max}\). Thus both scalings in Eq.~\eqref{eq:hierarchy} hold for the same
globally optimal design used to illustrate Eq.~\eqref{eq:general-design}.
The dashed guides in Fig.~\ref{fig:validation}(b) are shifted vertically for visibility;
the numerical curves show the unrescaled dimensionless quantities.

\subsection{Fast limit of the two-configuration reference}
Let \(D_\pi=\operatorname{diag}(\pi_N)\). We compute each gap as
the smallest positive eigenvalue of the symmetric matrix
\(-D_\pi^{-1/2}RD_\pi^{1/2}\). For an equally weighted auxiliary,
\begin{equation}
 \begin{gathered}
 g_0=g(R^{(0)}),\qquad
 \Reff=\frac{R^{(A)}+R^{(B)}}{2},\\
 g_{\rm aut}^{\infty}=g(\Reff).
 \end{gathered}
 \label{app:eq:CWReff}
\end{equation}
Here and in the two-configuration calculation below,
\(g_{\rm aut}^{\infty}\) is the gap of this specified mixture,
not the global optimum in Eq.~\eqref{app:eq:CWglobal-optimum}.
Its edgewise effective lowering is
\begin{equation}
 \Delta_m^{\rm eff}=\beta^{-1}\ln\!\left[
 \frac{e^{\beta\Delta_m^{(A)}}+e^{\beta\Delta_m^{(B)}}}{2}
 \right].
 \label{app:eq:CWeffective-barrier}
\end{equation}

\begin{figure}[!htbp]
\centering
\includegraphics[width=\linewidth]{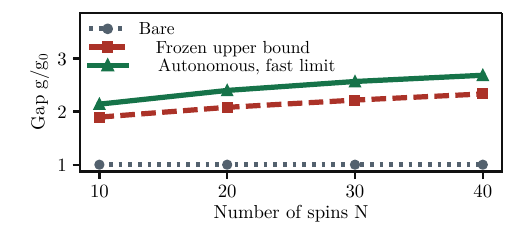}
\caption{\textbf{A simple two-configuration collective device.}
The equally weighted complementary profiles in
Eq.~\eqref{app:eq:CWdrop} have a fast-limit gap above the global
frozen upper bound at the same budget for each size.
This is the reference device used in Fig.~\ref{app:fig:CW}(d),
not the globally optimized $N$-configuration device in Fig.~\ref{fig:validation}. All gaps are normalized by the size-dependent bare gap.}
\label{app:fig:CWtwo}
\end{figure}

\begin{figure*}[!htbp]
\centering
\begin{minipage}[t]{0.48\textwidth}\centering
\includegraphics[width=\linewidth]{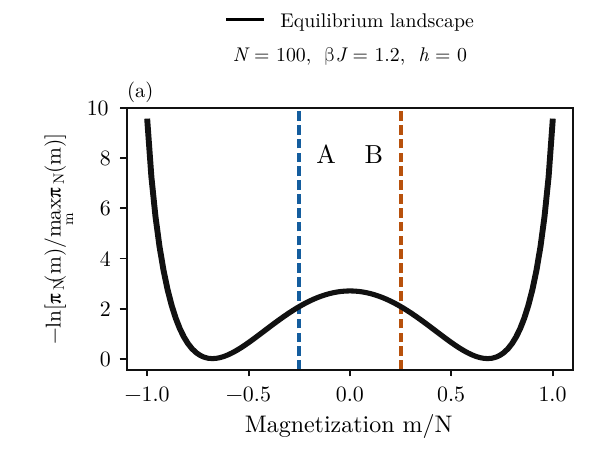}
\end{minipage}\hspace{0.012\textwidth}
\begin{minipage}[t]{0.48\textwidth}\centering
\includegraphics[width=\linewidth]{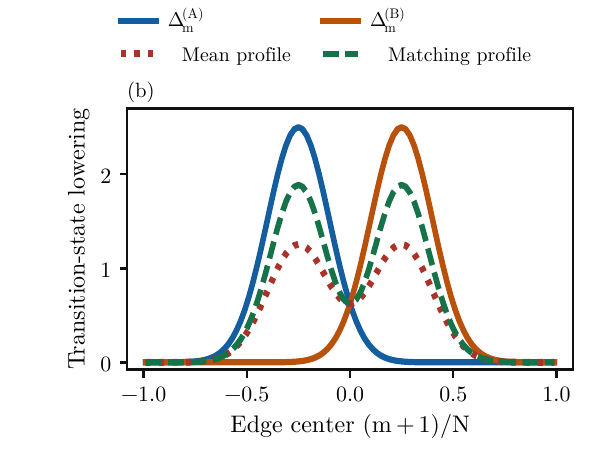}
\end{minipage}

\vspace{0.15em}
\begin{minipage}[t]{0.48\textwidth}\centering
\includegraphics[width=\linewidth]{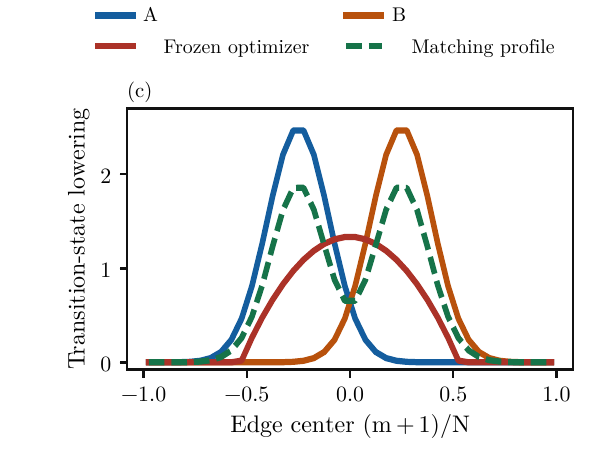}
\end{minipage}\hspace{0.012\textwidth}
\begin{minipage}[t]{0.48\textwidth}\centering
\includegraphics[width=\linewidth]{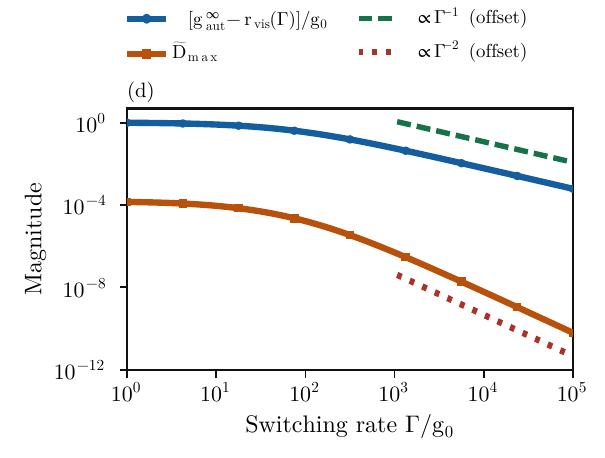}
\end{minipage}
\caption{\textbf{Two-configuration Curie--Weiss reference.}
(a) Shifted equilibrium magnetization landscape for \(N=100\).
(b) Complementary frozen profiles and their rate-averaged effective lowering.
(c) For \(N=40\), the numerical frozen optimizer is obtained over the full
simplex in Eq.~\eqref{app:eq:CWstaticclass}; the fast effective profile requires
a larger frozen budget. (d) Finite-switching rate deficit and numerically
sampled peak \(\widetilde D\) for \(N=40\).}
\label{app:fig:CW}
\end{figure*}

Figure~\ref{app:fig:CWtwo} and Table~\ref{app:tab:CW} show that this
mixture exceeds the same frozen upper bound by
\(12.9\%\)--\(15.8\%\), while remaining well below the global
autonomous optimum.
The equilibrium landscape in Fig.~\ref{app:fig:CW}(a) displays the
two magnetization basins at $N=100$.
Panel (b) resolves the complementary lowerings: each conditional
profile favors a different part of the path, while rate averaging
enhances both.
For $N=40$, the optimized frozen profile in panel (c) redistributes
the same budget over the full edge simplex; reproducing the effective
profile instead requires a larger frozen budget.

\begin{table*}[t]
\caption{Two-configuration reference compared with the frozen optimum
for the exactly lumped Curie--Weiss chain. Gaps are normalized by $g_0$;
$g_{\rm st}^{\rm ub}$ is the global upper bound in
Eq.~\eqref{app:eq:CWcertificate}, and $g_{\rm aut}^{\infty}$ refers only
to the equally weighted profiles in Eq.~\eqref{app:eq:CWdrop}.
The numerical optimum and the global upper bound differ by less
than $5\times10^{-7}g_0$ for every listed $N$; their displayed
values coincide after rounding.}
\label{app:tab:CW}
\centering
\begin{ruledtabular}
\begin{tabular}{c c c c c}
$N$ &
$g_{\rm st}^{\rm num}/g_0$ &
$g_{\rm st}^{\rm ub}/g_0$ &
$g_{\rm aut}^{\infty}/g_0$ &
$g_{\rm aut}^{\infty}/g_{\rm st}^{\rm ub}$ \\
\hline
10 & 1.895 & 1.895 & 2.14 & 1.129 \\
20 & 2.079 & 2.079 & 2.40 & 1.153 \\
30 & 2.215 & 2.215 & 2.57 & 1.158 \\
40 & 2.331 & 2.331 & 2.68 & 1.151 \\
\end{tabular}
\end{ruledtabular}
\end{table*}

\subsection{Finite switching in the two-configuration device}
At \(N=40\), vary only \(\Gamma\) for the same complementary
profiles and diagonalize the joint generator
\begin{equation}
 \begin{gathered}
 \dot P_{mc}=\sum_{m'}R_{mm'}^{(c)}P_{m'c}
 +\Gamma\sum_{c'}\widehat G_{cc'}P_{mc'},\\
 \widehat G=\begin{pmatrix}-1/2&1/2\\1/2&-1/2\end{pmatrix}.
 \end{gathered}
 \label{app:eq:CWjoint}
\end{equation}
Its equilibrium distribution is \(\Pi_{mc}=\pi_N(m)/2\).
Let \(\phi_2\) be the slow backward eigenfunction,
\(\Reff^{\mathsf T}\phi_2=-g_{\rm aut}^{\infty}\phi_2\), normalized by
\(\sum_m\pi_N(m)\phi_2(m)=0\) and
\(\sum_m\pi_N(m)\phi_2(m)^2=1\). In the symmetric representation,
we identify the corresponding joint decay rate \(r_{\rm vis}(\Gamma)\) by the
largest absolute eigenvector overlap with the embedded slow mode
\(u_{mc}=\sqrt{\pi_N(m)/2}\,\phi_2(m)\). This follows the
system-like branch approaching \(g_{\rm aut}^{\infty}\); at small \(\Gamma\),
it need not be the smallest nonzero joint decay rate.

For every switching rate, we use the same full-support product initial
state,
\begin{equation}
 \begin{gathered}
 p_m(0)=\pi_N(m)\left[1+
 \frac{0.2\,\phi_2(m)}{\max_{m'}|\phi_2(m')|}\right],\\
 P_{mc}(0)=\frac{p_m(0)}{2}.
 \end{gathered}
 \label{app:eq:CWinitial}
\end{equation}
Spectral propagation gives \(\Dt(t)\) from
Eq.~\eqref{app:eq:memory-def}. Its sampled peak uses 480 uniform
points on \(0\leq g_0t\leq12\) and 170 logarithmic points resolving
\(t=O(\Gamma^{-1})\), for 105 logarithmically spaced rates
\(1\leq\Gamma/g_0\leq10^5\).
Fits over \(10^4\leq\Gamma/g_0\leq10^5\) give
\begin{equation}
 \frac{g_{\rm aut}^{\infty}-r_{\rm vis}(\Gamma)}{g_0}
 \propto(\Gamma/g_0)^{-0.997},
 \qquad
 \Dt_{\max}\propto(\Gamma/g_0)^{-1.99}.
 \label{app:eq:CWfits}
\end{equation}
These results, shown in Fig.~\ref{app:fig:CW}(d), agree with the
rate-deficit and peak-departure scalings in Eq.~\eqref{eq:hierarchy}.


\section{Continuous-space realization and numerical validation}
\label{app:continuous}
\label{sec:realization}

This section specifies the continuous realization and its frozen benchmark.
The final example gives the \(41.1\%\) gain reported in the main text while
preserving minimum curvatures.

\subsection{Fixed two-dimensional landscape}
\begin{figure*}[t]
\centering
\includegraphics[height=0.26318957\textwidth]{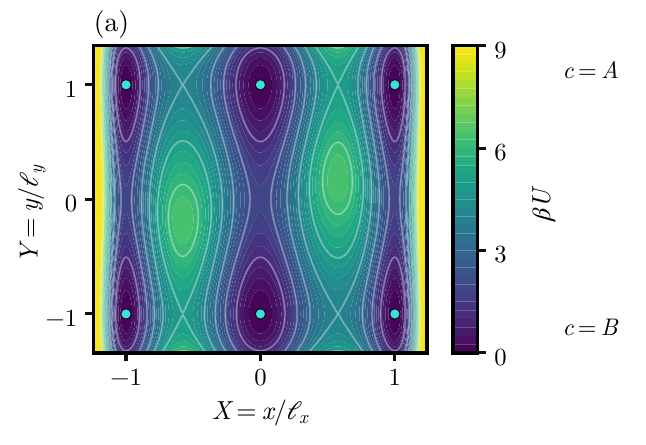}%
\hspace{0.012\textwidth}%
\includegraphics[height=0.26318957\textwidth]{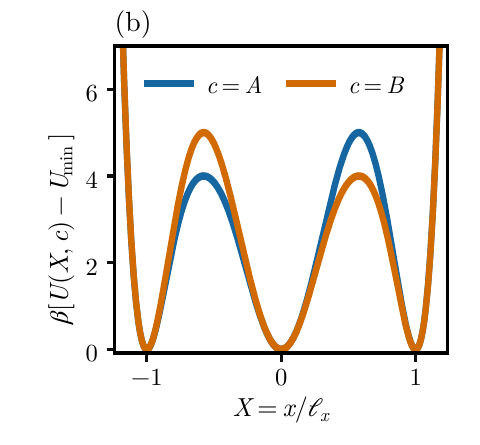}%
\hspace{0.012\textwidth}%
\includegraphics[height=0.26318957\textwidth]{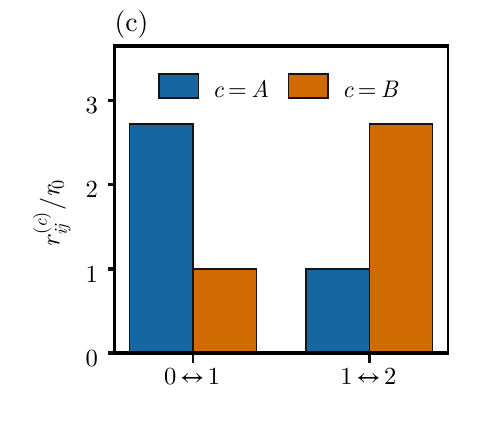}
\caption{\textbf{Passive time-independent colloidal landscape.}
(a) Fixed potential in Eq.~\eqref{app:eq:colloid-potential}. The six minima
encode \(i=0,1,2\) and \(c=A,B\). (b) Conditional cuts through the auxiliary
minima: \(A\) lowers the left system barrier and \(B\) the right barrier,
while leaving the locations and potential values of the three system minima
unchanged. The smooth coupling also changes local curvatures, which are kept in
the Fokker--Planck calculation. (c) Arrhenius estimates \(r_{ij}^{(c)}/r_0\),
where \(r_{ij}^{(c)}\) is the conditional rate on edge \(i\leftrightarrow j\)
and \(r_0\) is the uncoupled reference rate with the same prefactor.
Curvature changes and weak cross-talk are neglected in this panel.}
\label{app:fig:passive-landscape}
\end{figure*}

Let \(X=x/\ell_x\) and \(Y=y/\ell_y\) be dimensionless coordinates.  A
convenient fixed potential is
\begin{equation}
 \begin{aligned}
 \beta U(X,Y)
 & =u_S(X)+u_C(Y)\\
 &\quad-\delta\!\left[
 h_L(X)\chi_A(Y)+h_R(X)\chi_B(Y)
 \right],
 \end{aligned}
 \label{app:eq:colloid-potential}
\end{equation}
where
\begin{align}
 u_S(X)&=\frac{135}{4}X^2(X^2-1)^2,
 &
 u_C(Y)&=B_C(Y^2-1)^2,
 \label{app:eq:separable-potential}\\
 \chi_A(Y)&=\frac{1+\tanh(Y/\sigma_Y)}{2},
 &
 \chi_B(Y)&=1-\chi_A(Y),
 \label{app:eq:aux-selectors}
\end{align}
The complementary coupling coefficients define the nominal barrier
lowerings, in units of \(k_{\rm B}T\),
\begin{equation}
 \begin{gathered}
 \Delta_L(Y)=\delta\chi_A(Y),\qquad
 \Delta_R(Y)=\delta\chi_B(Y),\\
 \Delta_L(Y)+\Delta_R(Y)=\delta.
 \end{gathered}
 \label{app:eq:colloid-budget}
\end{equation}
Equation~\eqref{app:eq:colloid-budget} constrains coupling coefficients;
actual saddle positions and curvatures may also change. Each fixed $Y_0$
gives a member of the pinned comparison below. The balanced slice $Y_0=0$
is unstable in the autonomous double well, but can be stabilized in a
separate run by adding the transverse confinement
$u_C^{\rm pin}(Y)=\kappa(Y-Y_0)^2/2$ and taking large $\kappa$.
Thus the frozen comparison includes all allocations, not only the two
stable auxiliary conformations. We choose saddle-localized functions
\begin{align}
 h_L(X)
 &=
 \frac{X^2(X^2-1)^2}{4/27}
 \exp\!\left[
 -\frac{(X+1/\sqrt3)^2}{2\sigma_X^2}
 \right],
 \nonumber\\
 h_R(X)
 &=
 \frac{X^2(X^2-1)^2}{4/27}
 \exp\!\left[
 -\frac{(X-1/\sqrt3)^2}{2\sigma_X^2}
 \right].
 \label{app:eq:saddle-localized-coupling}
\end{align}
The coupling and its first derivatives vanish at $X=-1,0,1$, so all
six minimum positions and potential values remain unchanged. Local
curvatures can change, which is why we use the full Fokker--Planck
operator and include the curvature-preserving control below.
The uncoupled longitudinal barrier is $5\kB T$. With
$\sigma_X=0.22$, $\sigma_Y=0.25$, and $\delta=1$, configuration
$A$ ($Y\simeq+1$) lowers the left barrier to approximately $4\kB T$;
configuration $B$ acts on the right. At the opposite nominal saddle,
spatial cross-talk is $1.04\times10^{-6}$ and selector leakage is
$3.35\times10^{-4}$, giving a total lowering of approximately
$3.36\times10^{-4}\kB T$.
The parameter $B_C$ controls the transverse barrier; reflection symmetry
keeps the two auxiliary-well weights equal.
Figure~\ref{app:fig:passive-landscape}(a) identifies the three visible states
in each auxiliary well. Panel (b) shows complementary barrier lowerings at
$Y\simeq\pm1$ without shifting the system minima; panel (c) gives their
Arrhenius rate estimates, neglecting curvature changes and weak cross-talk
retained in the full Fokker--Planck calculation.

The overdamped particle dynamics is
\begin{align}
 \dot x&=-D_x\beta\,\partial_x U(x,y)
 +\sqrt{2D_x}\,\xi_x(t),
 \nonumber\\
 \dot y&=-D_y\beta\,\partial_y U(x,y)
 +\sqrt{2D_y}\,\xi_y(t),
 \label{app:eq:colloid-langevin}
\end{align}
with
\(\langle\xi_\mu(t)\xi_\nu(t')\rangle
=\delta_{\mu\nu}\delta(t-t')\).
Equation~\eqref{app:eq:colloid-langevin} obeys detailed balance with respect
to \(\rho_{\rm eq}(x,y)\propto e^{-\beta U(x,y)}\). Metastable basin dynamics
admits a reversible six-state approximation~\cite{SchutteEtAl2011}, but
curvature changes can prevent basin weights from factorizing exactly as
\(\pi_iq_c^*\) at finite parameters. The direct Fokker--Planck calculation
assumes neither factorization nor a six-state reduction. In the high-barrier, saddle-localized limit the
construction approaches the finite-state model, provided the intrabasin
partition factors also approach the product form discussed in
Appendix~\ref{app:averaging}.
Slow eigenmodes characterize metastable kinetics \cite{SchwantesEtAl2014},
while thermodynamically consistent coarse graining preserves local
detailed balance \cite{FalascoEsposito2021}.

The experiments cited in the main text provide static optical landscapes,
activated transitions, and single-particle readout. A representative scale is
\begin{equation}
 \begin{gathered}
 \ell_x=0.7\,\mu{\rm m},\qquad
 2\ell_y=0.7\,\mu{\rm m},\\
 D_x\simeq D_y\simeq0.5\,\mu{\rm m}^2{\rm s}^{-1},
 \end{gathered}
 \label{app:eq:representative-colloid-parameters}
\end{equation}
for a submicron bead in water.  The transverse barrier \(B_C\) is adjusted
between separate runs and kept fixed during each trajectory.  When a two-state Markov description is adequate, residence times define
an effective switching rate $k_{A\to B}+k_{B\to A}$.
Varying $B_C$ changes the landscape and is distinct from the fixed-device
limit in Eq.~\eqref{eq:hierarchy}. The mobility sweep below instead holds
$U$ fixed and varies the scaled transverse mobility $r$.

\subsection{Direct Fokker--Planck validation}
Writing \(u=\beta U\) and
\(\tau=D_xt/\ell_x^2\), Eq.~\eqref{app:eq:colloid-langevin} gives
\begin{equation}
 \begin{aligned}
 \partial_\tau\rho
 &=\partial_X\!\left[\partial_X\rho+(\partial_Xu)\rho\right]\\
 &\quad+r\,\partial_Y\!\left[\partial_Y\rho+(\partial_Yu)\rho\right],\\
 r&=\frac{D_y/\ell_y^2}{D_x/\ell_x^2}.
 \end{aligned}
 \label{app:eq:fokker-planck}
\end{equation}
The representative lengths and diffusivities in
Eq.~\eqref{app:eq:representative-colloid-parameters} give \(r=4\). Thus
\(r\) is a scaled transverse mobility, not simply \(D_y/D_x\).

Pinning \(Y=Y_0\) assigns the fraction \(\lambda=\chi_A(Y_0)\) of the
coupling to the left barrier. Since \(u_C(Y_0)\) is constant along the
longitudinal coordinate, the frozen potential is
\begin{equation}
 \begin{aligned}
 u_\lambda(X)
 &=u_S(X)\\
 &\quad-\delta\!\left[\lambda h_L(X)+(1-\lambda)h_R(X)\right],\\
 &\qquad 0\leq\lambda\leq1.
 \end{aligned}
 \label{app:eq:pinned-family}
\end{equation}
This family includes the balanced profile \(\lambda=1/2\) and the limiting
allocations \(\lambda=0,1\); finite \(Y_0\) gives \(0<\lambda<1\).
Using the spectral-gap definition above, we write \(g_{\rm aut}(r)\)
for the full two-dimensional gap and \(g_{\rm st}(\lambda)\) for the frozen
one-dimensional gap, with
\(g_{\rm st}^{\rm opt}=\max_{0\leq\lambda\leq1}g_{\rm st}(\lambda)\).
Rates are expressed in units of \(D_x/\ell_x^2\).
The computed curve is symmetric under \(\lambda\mapsto1-\lambda\); a
continuous bounded optimization gives \(\lambda_{\rm opt}\simeq0.5\),
confirming the balanced maximum.

\subsection{Discrete operator and frozen-profile bound in continuous space}
For uniform grid spacing \(h_\mu\) and dimensionless mobility \(d_\mu\)
in direction \(\mu\), set \(z=u_\beta-u_\alpha\) for neighboring cells
\(\alpha,\beta\). Integrating
\(J_\mu=-d_\mu e^{-u}\partial_\mu(e^u\rho)\), with a constant
edge flux and a linear interpolation of \(u\), gives
\begin{equation}
 J_{\alpha\to\beta}
 =\frac{d_\mu}{h_\mu}
 [\mathcal B(z)\rho_\alpha-\mathcal B(-z)\rho_\beta],
 \label{app:eq:sg-flux}
\end{equation}
where \(\mathcal B(z)=z/(e^z-1)\) and \(\mathcal B(0)=1\).
Dividing by the cell width gives the Scharfetter--Gummel rates
\begin{equation}
 k_{\beta\alpha}^{(\mu)}
 =\frac{d_\mu}{h_\mu^2}\mathcal B(u_\beta-u_\alpha).
 \label{app:eq:sg-rates}
\end{equation}
They satisfy
\(k_{\beta\alpha}^{(\mu)}e^{-u_\alpha}
=k_{\alpha\beta}^{(\mu)}e^{-u_\beta}\), so the discrete chain
is reversible with the sampled Boltzmann weights. Reflecting boundaries
are imposed by omitting transitions across the outer faces.
Preserving equilibrium structure is also central to other
thermodynamically consistent solvers \cite{Holubec2019}.

On a fixed longitudinal grid, consider the frozen family
\(u_\lambda(X_i)=u_0(X_i)+\lambda v_i\), with
\(0\leq\lambda\leq1\), and choose time units such that \(d_X=1\).
The unnormalized weights are \(m_i(\lambda)=e^{-u_\lambda(X_i)}\),
and the edge conductances are
\begin{equation}
 c_{i,i+1}(\lambda)
 =\left[h_X^2\int_0^1
 e^{(1-s)u_\lambda(X_i)+s u_\lambda(X_{i+1})}\,ds\right]^{-1}.
 \label{app:eq:sg-conductance}
\end{equation}
Their common normalization cancels from the gap formula:
\begin{equation}
 g_h(\lambda)
 =\min_{f\ \mathrm{nonconstant}}
 \frac{\sum_i c_{i,i+1}(\lambda)(f_{i+1}-f_i)^2}
 {\min_a\sum_i m_i(\lambda)(f_i-a)^2}.
 \label{app:eq:grid-gap-quotient}
\end{equation}
For a sampled parameter \(\lambda_j\), set
\(d_i=(\lambda-\lambda_j)v_i\), \(d_- =\min_i d_i\), and
\(d_+=\max_i d_i\). Both the weights and conductances at
\(\lambda\), divided by their values at \(\lambda_j\), lie
between \(e^{-d_+}\) and \(e^{-d_-}\). Thus the numerator in
Eq.~\eqref{app:eq:grid-gap-quotient} is bounded above by
\(e^{-d_-}\) times its reference value, while the denominator is
bounded below by \(e^{-d_+}\) times its reference value.
Evaluating the quotient on a minimizing function at \(\lambda_j\)
therefore gives
\begin{equation}
 g_h(\lambda)
 \leq e^{\omega_h|\lambda-\lambda_j|}g_h(\lambda_j),
 \label{app:eq:grid-parameter-bound}
\end{equation}
where \(\omega_h=\max_i v_i-\min_i v_i\).
If every parameter value lies within \(\rho_\lambda\) of a sample,
then
\begin{equation}
 \max_{0\leq\lambda\leq1}g_h(\lambda)
 \leq e^{\omega_h\rho_\lambda}\max_j g_h(\lambda_j).
 \label{app:eq:grid-frozen-bound}
\end{equation}
For sample spacing \(1/1000\), including both endpoints,
\(\rho_\lambda=1/2000\). This bounds the unsampled frozen profiles
on each spatial grid; it does not estimate the continuum discretization
error.

\subsection{Spectral validation and grid convergence}
Reflecting boundaries are imposed on $X\in[-1.4,1.4]$ and
$Y\in[-1.8,1.8]$. We use the reversible Scharfetter--Gummel operator
in Eq.~\eqref{app:eq:sg-rates}~\cite{ScharfetterGummel1969}.
On an independent
\(61\times81\) audit grid, the maximum relative detailed-balance residual is
\(6.4\times10^{-16}\). Shift-invert and smallest-magnitude settings of the same ARPACK
eigensolver agree on the gap within
\(5\times10^{-13}\), and all reported eigenpair residuals are below
\(2.1\times10^{-11}\).

\begin{figure}[t]
\centering
\includegraphics[width=\linewidth]{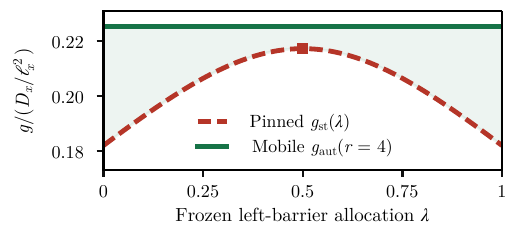}
\caption{\textbf{Mobile and pinned relaxation in the smooth potential.}
The pinned gap is optimized over the full allocation interval
\(0\leq\lambda\leq1\) in Eq.~\eqref{app:eq:pinned-family}.
The full two-dimensional mobile gap at \(r=4\) exceeds its maximum
by approximately \(3.7\%\) on the common longitudinal grid, with
\(n_X=221\) and \(n_Y=241\) for the mobile calculation.
The comparison retains all coordinate modes; it does not use a
six-state approximation.}
\label{app:fig:continuous-comparison}
\end{figure}

\begin{figure*}[t]
\centering
\begin{minipage}[t]{0.31\textwidth}
 \centering
 \includegraphics[width=\linewidth]{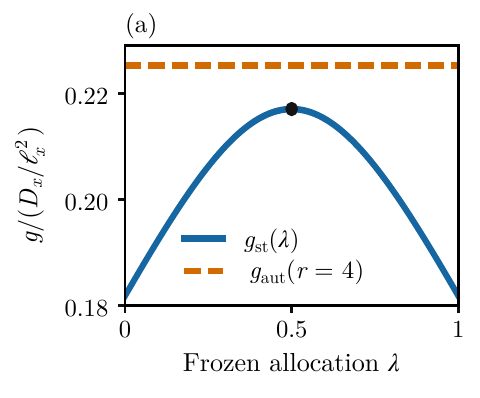}
\end{minipage}\hspace{0.012\textwidth}
\begin{minipage}[t]{0.31\textwidth}
 \centering
 \includegraphics[width=\linewidth]{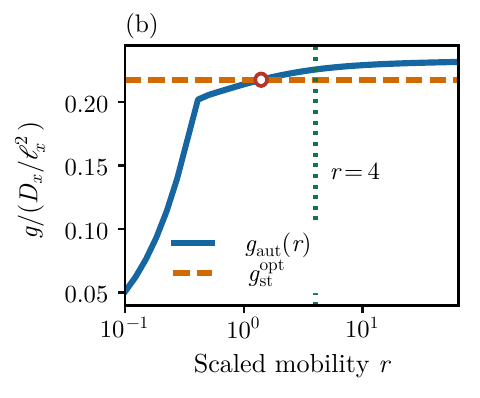}
\end{minipage}\hspace{0.012\textwidth}
\begin{minipage}[t]{0.31\textwidth}
 \centering
 \includegraphics[width=\linewidth]{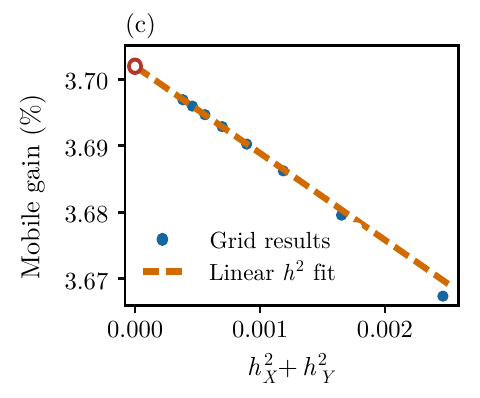}
\end{minipage}
\caption{\textbf{Direct Fokker--Planck validation of the smooth landscape.}
(a) The gap of every pinned profile in
Eq.~\eqref{app:eq:pinned-family}, computed on 601 longitudinal cells;
the balanced profile is optimal, while the full mobile gap at \(r=4\)
on the \(201\times221\) grid lies above it. (b) Full two-dimensional gap versus
scaled transverse mobility. The open circle marks \(r_c\simeq1.4\), and the
vertical dotted line marks the representative value \(r=4\). The bend marks
the crossover from auxiliary-limited to system-like relaxation. (c) Grid
convergence of the mobile advantage;
the linear \(h^2\) extrapolation gives \(3.702\protect\%\).}
\label{app:fig:fp-validation}
\end{figure*}

For \(B_C=1.8\), \(\delta=1\), \(\sigma_X=0.22\), and
\(\sigma_Y=0.25\), Figs.~\ref{app:fig:continuous-comparison} and
\ref{app:fig:fp-validation}(a) compare the mobile gap with the pinned
family, whose maximum occurs at \(\lambda=1/2\). At \(r=4\), a
linear extrapolation in \(h_X^2+h_Y^2\) over the five finest grids gives
\begin{equation}
 \begin{gathered}
 g_{\rm aut}(r=4)\simeq0.225\,\frac{D_x}{\ell_x^2},\\
 g_{\rm st}^{\rm opt}\simeq0.217\,\frac{D_x}{\ell_x^2},\\
 \frac{g_{\rm aut}(r=4)}{g_{\rm st}^{\rm opt}}=1.03702.
 \end{gathered}
 \label{app:eq:fp-result}
\end{equation}
In Fig.~\ref{app:fig:continuous-comparison}, the pinned scan and mobile gap use the
same longitudinal grid, \(n_X=221\), with \(n_Y=241\) for the mobile
calculation. Their dimensionless gaps are
\(g_{\rm st}^{\rm opt}\ell_x^2/D_x\simeq0.217\) and
\(g_{\rm aut}(r=4)\ell_x^2/D_x\simeq0.225\), giving an advantage of
approximately \(3.7\%\). This finite-grid comparison is distinct from, and
consistent with, the extrapolated values in Eq.~\eqref{app:eq:fp-result}.
The joint gap crosses the optimal pinned value at
\(r_c\simeq1.4\) on the production grid
[Fig.~\ref{app:fig:fp-validation}(b)]. The ratios on
\(121\times141\), \(161\times181\), and \(201\times221\) grids are
\(1.03686\), \(1.03693\), and \(1.03696\), respectively, and the
continuum extrapolation is shown in Fig.~\ref{app:fig:fp-validation}(c).
The equilibrium mass in the outermost cell layer is
\(1.9\times10^{-6}\), and enlarging the computational box changes the
\(201\times221\)-grid ratio by \(5.1\times10^{-7}\). Using the nominal separatrices
\(X=\pm1/\sqrt3\) and \(Y=0\), the largest relative deviation of the six
basin weights from the product of their marginals is \(2.3\%\); the visible
three-basin weights differ from those of the balanced pinned landscape by at
most \(1.3\times10^{-5}\). No Kramers prefactor approximation or basin coarse
graining enters this calculation. The gain is positive for \(r>r_c\) in the scanned range; it is negative
at \(r=1\). Finite transverse mobility, curvature changes, and the full
smooth barrier geometry are all retained. This is a comparison within the
specified potential family, not an application of the finite-state
optimality theorem to arbitrary continuous potentials.

\subsection{Control with unchanged minimum curvatures}
To test whether the advantage requires changing local well curvatures,
replace the coupling functions in
Eq.~\eqref{app:eq:saddle-localized-coupling} by
\begin{equation}
 h_{L,R}^{(2)}(X)=
 \left[\frac{X^2(X^2-1)^2}{4/27}\right]^2
 \exp\!\left[-\frac{(X\pm1/\sqrt3)^2}{2\sigma_X^2}\right],
 \label{app:eq:curvature-preserving-coupling}
\end{equation}
with the upper sign for the left saddle. The envelope now has fourth-order
zeros at \(X=-1,0,1\). Thus \(h^{(2)}\), its first derivative, and its
second derivative vanish at every system minimum. The six minimum values,
positions, and Hessians are exactly those of the uncoupled potential. At
\(B_C=1.8\), the dimensionless Hessian is
\(\operatorname{diag}(270,14.4)\) in the side wells and
\(\operatorname{diag}(67.5,14.4)\) in the central wells. Anharmonic basin
partition factors can still change, so this control does not impose exact
factorization of the six basin probabilities.

All other parameters are unchanged. We assemble the reversible finite-volume
operator for each grid and sample the pinned family at
\(\lambda_j=j/1000\), \(j=0,\ldots,1000\), followed by a local refinement
near its largest sampled value. Every grid gives the largest pinned gap at
\(\lambda\simeq1/2\). Table~\ref{app:tab:curvature-control} shows an
advantage of approximately \(2.78\%\) on the finest grid; the ratio changes by
\(1.50\times10^{-5}\) between the two finest grids. The largest mobile
and pinned eigenpair residuals are \(1.83\times10^{-11}\) and
\(3.07\times10^{-12}\), respectively.

\begin{table*}[t]
\caption{Continuous-space control with the coupling in
Eq.~\eqref{app:eq:curvature-preserving-coupling}. Rates are in units of
\(D_x/\ell_x^2\), and \(r=4\). The last column compares the mobile gap
with the upper bound covering the entire pinned interval on that spatial
grid.}
\label{app:tab:curvature-control}
\centering
\begin{ruledtabular}
\begin{tabular}{c c c c c}
Grid & \(g_{\rm aut}\) & \(g_{\rm st}^{\rm num}\) &
\(g_{\rm aut}/g_{\rm st}^{\rm num}\) &
\(g_{\rm aut}/g_{\rm st}^{\rm ub}\)\\
\hline
\(101\times121\)&0.219457&0.213555&1.02764&1.02660\\
\(141\times161\)&0.219084&0.213178&1.02771&1.02667\\
\(181\times201\)&0.218930&0.213021&1.02774&1.02670\\
\(221\times241\)&0.218851&0.212942&1.02775&1.02672\\
\end{tabular}
\end{ruledtabular}
\end{table*}

Equation~\eqref{app:eq:grid-frozen-bound} bounds all unsampled
allocations. Including a \(10^{-9}\) absolute numerical allowance on
sampled gaps gives \(g_{\rm st}^{\rm ub}\leq0.213155\) on the finest
grid. The mobile gap exceeds this bound by approximately \(2.67\%\).
This bounds each discrete operator; Table~\ref{app:tab:curvature-control}
separately assesses spatial error. The advantage thus persists with
unchanged minimum curvatures.

\subsection{Larger gain with unchanged minimum curvatures}
The gains above are not an upper limit for this potential family.
Retain Eq.~\eqref{app:eq:curvature-preserving-coupling}, but set
\(u_S(X)=54X^2(X^2-1)^2\), giving a bare longitudinal barrier of
\(8\kB T\), with \(\delta=2.5\) and \(\sigma_X=0.30\).
Keep \(B_C=1.8\), \(\sigma_Y=0.25\), and \(r=4\).
The mobile potential and its entire pinned family share these parameters;
the frozen optimum is recomputed over \(0\leq\lambda\leq1\).
This is a separate parameter set from the smaller-gain examples in
Figs.~\ref{app:fig:continuous-comparison} and~\ref{app:fig:fp-validation}
and Table~\ref{app:tab:curvature-control}.

The coupling leaves the six minimum positions, values, and Hessians
unchanged relative to the uncoupled potential at the same bare barrier
\(8\kB T\). The Hessians are now
\(\operatorname{diag}(432,14.4)\) in the side wells and
\(\operatorname{diag}(108,14.4)\) in the central wells.
A multistart stationary-point search finds six minima, seven saddles,
and two maxima. A separate scan of 1001 pinned allocations on
\(X\in[-4,4]\) finds three minima and two maxima for each allocation;
no additional wells are found in these numerical checks.

\begin{table*}[!htbp]
\caption{Larger continuous-space gain with unchanged minimum curvatures.
The parameters are \(u_S(X)=54X^2(X^2-1)^2\), \(\delta=2.5\),
\(\sigma_X=0.30\), \(B_C=1.8\), \(\sigma_Y=0.25\), and \(r=4\).
Rates are in units of \(D_x/\ell_x^2\).
The last column uses the upper bound over every pinned allocation
on the indicated spatial grid.}
\label{app:tab:enhanced-curvature-control}
\centering
\begin{ruledtabular}
\begin{tabular}{c c c c c}
Grid & \(g_{\rm aut}\) & \(g_{\rm st}^{\rm num}\) &
\(g_{\rm aut}/g_{\rm st}^{\rm num}\) &
\(g_{\rm aut}/g_{\rm st}^{\rm ub}\)\\
\hline
\(221\times241\)&0.0491657&0.0348546&1.41059&1.39885\\
\(301\times321\)&0.0491449&0.0348357&1.41076&1.39890\\
\(401\times421\)&0.0491343&0.0348261&1.41085&1.39891\\
\end{tabular}
\end{ruledtabular}
\end{table*}

Using the same reversible discretization and reflecting box as above,
Table~\ref{app:tab:enhanced-curvature-control} gives a gain of
approximately \(41.1\%\) on the finest grid. The pinned scan and local refinement
again locate the maximum at \(\lambda\simeq1/2\).
With 1001 equally spaced allocation samples and a \(10^{-9}\) allowance
on sampled gaps, Eq.~\eqref{app:eq:grid-frozen-bound} gives
\(g_{\rm st}^{\rm ub}\leq0.0351232\) on that grid. The mobile gap
exceeds this bound by approximately \(39.9\%\). This bound covers the allocation interval
for the discrete operator; spatial error is assessed separately.
The ratio changes by \(8.53\times10^{-5}\) between the two finest
grids. Expanding the box to \([-1.6,1.6]\times[-2.1,2.1]\) on a
\(345\times375\) grid, with nearly the same spacing as the
\(301\times321\) grid, changes the ratio by \(1.04\times10^{-6}\).
The largest mobile and pinned eigenpair residuals are below
\(4.6\times10^{-11}\) and \(1.2\times10^{-11}\), respectively.

Using the same nominal basin boundaries as above, converged quadrature
gives a maximum absolute difference of \(2.07\times10^{-5}\) between
the mobile and optimal pinned visible basin weights. The six-basin
distribution differs from the product of its marginals by at most
\(1.16\%\) in relative terms. Thus the larger gain retains close,
but not exact, equilibrium matching. It is measured against this
parameter set's own optimal pinned reference; it does not imply a
larger absolute relaxation rate than in the original lower-barrier
example. The numerical data and reproduction script accompany the
manuscript.

\subsection{Experimental observables and reversibility checks}
Compare mobile and pinned trajectories at the same bare longitudinal
potential, coupling amplitude, and mobility. Scan the transverse trap
center between runs and increase its stiffness until the inferred
longitudinal gap converges to the pinned value. Equilibrium fluctuation
measurements calibrate trap stiffness and mobility
\cite{BergSorensenFlyvbjerg2004}; the additional confinement remains
static during each relaxation trajectory.
Infer the gap from joint transition counts or mode-resolved populations:
visible observables can miss auxiliary modes, and relative entropy
decays at twice the contributing population rate near equilibrium.
Check metastability, cross-talk, diagonal transitions, and detailed balance
$k_{\alpha\beta}\Pi_\beta\simeq k_{\beta\alpha}\Pi_\alpha$ within
uncertainty. Optical forces can have nonconservative components
\cite{RoichmanEtAl2008}; stationary circulation or a reconstructed
nonconservative force would place an implementation outside the
reversible model. The coupled six-well experiment remains a proposal.

\renewcommand{\bibfont}{\small}
\setlength{\bibsep}{0pt}

\hypersetup{urlcolor=black}

\end{document}